%% file: main.tex
\documentclass{IEEEtran}

\usepackage{balance}  
\usepackage{graphicx} 
\usepackage{times}    
\usepackage{url}      
\usepackage{tabularx}
\usepackage[safe]{tipa}
\usepackage{amsmath}
\usepackage{amssymb}
\usepackage[boxruled]{algorithm2e}
\usepackage{algorithmic}
\usepackage{subfigure}
\usepackage{multirow}
\newcommand{\eat}[1]{}
\usepackage[font=small,skip=0pt]{caption}
\makeatletter
\def\url@leostyle{%
  \@ifundefined{selectfont}{\def\UrlFont{\sf}}{\def\UrlFont{\small\bf\ttfamily}}}
\makeatother
\newtheorem{theorem}{Theorem}[section]

\newtheorem{lemma}[theorem]{Lemma}
\newtheorem{proposition}[theorem]{Proposition}
\newtheorem{corollary}[theorem]{Corollary}

\newenvironment{definition}[1][Definition]{\begin{trivlist}
\item[\hskip \labelsep {\bfseries #1}]}{\end{trivlist}}

\eat{
\usepackage{amsmath,amsfonts,amssymb}

\newtheorem{theorem}{Theorem}[section]

}

\def\bdd{\text{\texthtd} }
\def\pprw{8.5in}
\def\pprh{11in}

\usepackage[pdftex,bookmarks=false]{hyperref}
\hypersetup{
pdfauthor={LaTeX},
pdfstartview={FitH},
colorlinks,
citecolor=black,
filecolor=black,
linkcolor=black,
urlcolor=black,
breaklinks=true,
}

\begin{document}

\title{Bounded Quadrant System: Error-bounded Trajectory Compression on
  the Go}

\author{
 $^\dag$Jiajun Liu \hspace{0.8cm} $^\dag$Kun Zhao \hspace{0.8cm}
 $^\dag$Philipp Sommer \hspace{0.8cm} \\
 $^*$Shuo Shang \hspace{0.8cm} $^\dag$Brano Kusy \hspace{0.8cm}
 $^\dag$Raja Jurdak\\
{$^\dag$AS Program, CSIRO, Pullenvale, Australia}\\
{ \{jiajun.liu, kun.zhao, philipp.sommer, raja.jurdak, brano.kusy\}@csiro.au}\\
{$^*$China University of Petroleum, Beijing, China}\\
{sshang@cup.edu.cn}
}

\maketitle

\begin{abstract}
Long-term location tracking, where trajectory compression is commonly
used, has gained high interest for many
applications in transport, ecology, and wearable computing. However, state-of-the-art
compression methods involve high space-time complexity or
achieve unsatisfactory compression rate, leading to rapid
exhaustion of memory, computation, storage and energy resources. 
We propose a novel online algorithm for error-bounded trajectory
compression called the Bounded Quadrant System (BQS), which compresses
trajectories with extremely small costs in space and time using
convex-hulls. In this algorithm, we build a virtual coordinate system
centered at a start point, and establish a rectangular bounding box as
well as two bounding lines in each of its quadrants. In each quadrant, 
the points to be assessed are bounded by the convex-hull formed by the
box and lines. Various compression error-bounds are therefore derived to 
quickly draw compression decisions without expensive error computations. 
In addition, we also propose a light version of the BQS version that
achieves $\mathcal{O}(1)$ complexity in both time and space for
processing each point to suit the most constrained computation
environments. Furthermore, we briefly demonstrate how this algorithm 
can be naturally extended to the 3-D case.

Using empirical GPS traces from flying foxes, cars 
and simulation, we
demonstrate the effectiveness of our algorithm in significantly
reducing the time and space complexity of trajectory compression, while
greatly improving the compression rates of the state-of-the-art
algorithms (up to 47\%). We then show that with this
algorithm, the operational time of the target 
resource-constrained hardware platform 
can be prolonged by up to 41\%.

\end{abstract}
\eat{
\keywords{
	Guides; instructions; author's kit; conference publications;
	keywords should be separated by a semi-colon.
}

\category{H.5.m.}{Information Interfaces and Presentation (e.g. HCI)}{Miscellaneous}

See: \url{http://www.acm.org/about/class/1998/}
for more information and the full list of ACM classifiers
and descriptors. 
}
 

\section{Introduction}
\input{intro.tex}

\section{Related Work}
\input{relatedwork.tex}

\section{Background}
\label{sec:mot}
In this section, we present the background of the study. The hardware system architecture used in the real-life bat tracking application is described. We also briefly introduce two existing solutions \cite{Heckbert95surveyof}\cite{opening_window}, for two reasons. First, by analyzing the existing algorithms we provide insights about why we need a better algorithm. Second, these two algorithms will be evaluated along with the proposed algorithm in a comparative study in the experiments.

\input{data.tex}

\section{Preliminaries}
\input{preliminaries.tex}

\section{Trajectory Compression with Bounded Quadrant System}
\subsection{Overview}
\input{overview.tex}


\input{compression.tex}

\subsection{Achieving Constant Time and Space Complexity}
\input{compression_fast.tex}

\input{generalization.tex}



\input{experiments.tex}

\section{Conclusion}
In this paper we present an online trajectory
compression algorithm for resourced-constrained environments. 
We first propose a convex-hull bounding structure 
called the Bounded Quadrant System, and then show tight bounds 
can be derived from it so that compression decisions will be 
efficiently determined without actual deviation calculations.

To further reduce the time and space complexity of the BQS algorithm,
a fast version of the BQS compression algorithm is also proposed. In
this version, error calculations are completely eliminated. Instead,
when uncertain of the error, the fast algorithm aggressively takes a
point. However due to the tight error bounded provided by the BQS,
the overhead in the compression rate is minimum, making it a
light-weight, efficient and effective algorithm, which is ideal for
 constrained computation environments.

As we establish the BQS algorithm, a discussion is also provided 
for the generalization and extensibility of
the BQS algorithm for the 3-D space as well as for a different error
metric. We have showed that such extensions are natural and
straightforward. BQS' flexibility to generalize to other applications
and settings is demonstrated.

To evaluate BQS, we have collected empirical data using a low-energy 
tracking platform called Camazotz on both animals and vehicles. 
To further widen the data variety, we also used synthetic dataset that 
is statistically representative of flying foxes' movement dynamics. 
In the experiments we evaluate the framework from various aspects.
We examine the pruning power of the original BQS algorithm,
demonstrating that the great pruning power of BQS leaves an ideal
opportunity for FBQS to exploit so that further improvement on
the time and space efficiencies is achieved without sacrificing much
compression rate. We present the actual compression rates of
BQS and fast BQS, and compare them to the results of competitive
methods. Comparison of the estimated operational time with different
algorithms is also presented. The actual run time is also reported.

There are a few immediate extensions to this work. The excellent
performance of the BQS algorithms provides a unique opportunity to
develop online and individualized smart systems for long-term tracking. For
instance, merging and ageing can be used on the historical trajectory
data to further reduce storage space. Individualized trajectory
and waypoint discovery can also be used to facilitate advanced applications
like real-time trip prediction or trip-duration estimation. Exploring
the potential of a 4-D BQS could be another interesting extension to
this work.


\bibliographystyle{IEEEtran}
\bibliography{main}

\section{Appendix: Discussion and Proof of Theorems}
\label{sec:app}
First we explain the purpose of splitting the space into four
quadrants and the properties obtained by this setup.

\begin{itemize}
\item No two bounding lines will intersect with the same edge of a
  bounding box, and every edge will have exactly
 one intersection with the bounding lines, except at the corner points
 or on the axes.
\item The angle between $l_{s,e}$ and either of the two bounding lines will be smaller than
  $90^\circ$, 
\item Hence, any point will be bounded by the convex hull formed by the
  points $\{l_1, l_2, u_1, u_2, c_n, c_f\}$. No bounding convex-hulls from two adjacent BQS will overlap. 
\end{itemize}

\begin{figure}[htp]
\vspace{-0.3cm}
\hspace{1.8cm}
\includegraphics[height=2.5cm]{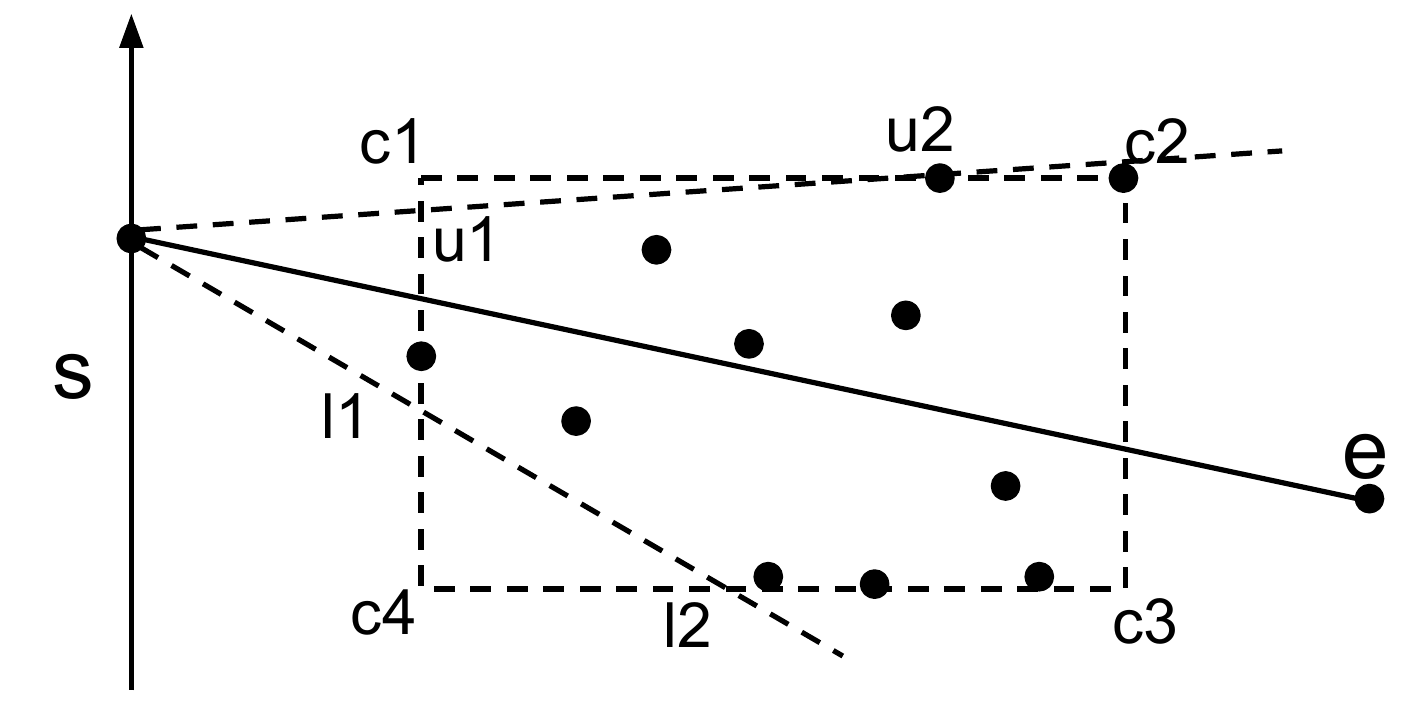}
\caption{Bounding system across quadrants}
\label{fig:ebqs}
\vspace{-0.5cm}
\end{figure}

Splitting the space into four quadrants ensures that the
aforementioned properties hold. Otherwise, the only case that the convex hull
  $\overline{l_1 l_2 c_f u_2 u_1 c_n}$ may not cover all the
  points is when there is an edge with zero intersection from the
  bounding lines, as depicted in Figure \ref{fig:ebqs}. In this case,
  $c_f$ is  $c_3$, however $c_2$ has a greater deviation to $l_{s,e}$
  than any point in $\{l_1, l_2, u_1, u_2, c_n, c_f\}$. Notice
  that when the system is in a single quadrant, the layout would not
  be a possibility.

Next we present the proof for Theorems
\ref{thm:1}, \ref{thm:2}, and \ref{thm:4}:
\paragraph{Proof of Theorem \ref{thm:1}} Using the first quadrant as
example as shown in Figure \ref{fig:bqs}, we
can state there are no points in areas $\overline{c_1u_1u_2}$ or
$\overline{c_3l_1l_2}$. Now we have the following properties:
\begin{itemize}
\item $d(u_1, l_{s,e}) \leq d^{max}(p, l_{s,e}) \leq d( u_2, l_{s,e} )$ :
  Because the angle between $l_{s,e}$ and $l_{u_1,u_2}$ is less than
  or equal to 90$^\circ$, if we extend $\overline{u_2c_2}$ to
  intersect with $l_{s,e}$ at $p_1$, the three vertices form a
  triangle $\overline{u_2p_1s}$, so the greatest distance from
  any point in the triangle to the edge $l_{s,e}$ would be $d(u_2,
  l_{s,e})$. Then because the bounding line dictates that there must
  be at least one point (denoted as $p^{u}$) in the line segment $\overline{u_1, u_2}$, which does
  not intersect with line $l_{s,e}$, we have $d(p^{u}, l_{s,e}) \leq
  min\{ d(u_1, l_{s,e}), d(u_2, l_{s,e}) \}$. In this case we have
  $min\{ d(u_1, l_{s,e}), d(u_2, l_{s,e}) \}=d(u_1, l_{s,e})$, hence we
  have $d(u_1, l_{s,e}) \leq d^{max}(p_i, l_{s,e}) \leq d( u_2, l_{s,e} )$.
\item $d(l_1, l_{s,e}) \leq d^{max}(p_i, l_{s,e}) \leq d( l_2, l_{s,e}
  )$ : proof is similar as above using the triangle $\overline{u_2p_1s}$. We could also see
  that by using $\overline{l_2p_1s}$ we have two triangles that
  together contain the convex hull
  $\overline{c_4u_1u_2c_2l_2l_1}$ which bounds the points, so we have $d^{max}(p_i, l_{s,e}) \leq  max\{
  d^{intersection} \}$ as the upper bound. Similarly, we can have $max\{ d(u_1, l_{s,e}),
  d(l_1, l_{s,e}) \}  \leq d^{max}(p_i, l_{s,e}) $ as the lower
  bound.

\item $d^{max}(p, l_{s,e}) \geq max(d^{corner-nf})$: This property is based on the fact that
  the line $l_{s,e}$ will not intersect with both edges that a corner
  point is on except at the corner points, while every edge must have at least one point on it.
  In this quadrant, $c_n$ is $c_4$ and $c_f$ is $c_2$. Because there is at least one point on
  the edge $\overline{c_1c_2}$, and $c_2$ is the closest point to
  $l_{s,e}$ on $\overline{c_1c_2}$, we have $d^{max}(p, l_{s,e}) \geq d(c_2,
  l_{s,e})$. With an identical case of edge $\overline{c_1c_4}$, we
  have $d^{max}(p, l_{s,e}) \geq max(d^{corner-nf})$ when we combine the
  lower bounds.
\end{itemize}

The line $l_{s,e}$ may intersect with the bounding box in different
angle and at different locations but with the properties guaranteed
by the BQS, we can use the same proof for all cases.

Theorems \ref{thm:2} is proven with similar techniques.
Theorem \ref{thm:4}, i.e. cases in which the line $l_{s,e}$ is in a different quadrant from
the bounding box, can
be proven using the same proof as for Theorem \ref{thm:0}.

\end{document}

%% file: intro.tex
Location tracking is increasingly important for transport, ecology, and wearable computing. In particular, long-term tracking of spatially spread assets, such as wildlife~\cite{Jurdak_tosn13}, augmented reality glasses~\cite{van2010survey}, or bicycles~\cite{DBLP:journals/tosn/EisenmanMLPAC09} provides high resolution trajectory information for better management and services.  For smaller moving entities, such as flying foxes~\cite{DBLP:conf/ipsn/JurdakSKKCMW13} or pigeons~\cite{nagy2010hierarchical}, the size and weight of tracking devices are constrained, which presents the challenge of obtaining detailed trajectory information subject to memory, processing, energy and storage constraints. 

Consider tracking of flying foxes as a motivating scenario. The computing platform is inaccessible once deployed and it is constrained in computational resources. The position data, on the other hand, is acquired
in a stream fashion. The RAM available on the platform barely reaches 4 KBytes, while the storage space is only 1 MB to store trajectories over weeks and months before they can be offloaded. The combination of long-term operational requirement and constrained resources therefore 
requires an intelligent online algorithm that can process the incoming points instantaneously, i.e in constant space and time, and that can achieve high compression rate. 

Current trajectory compression algorithms often fail to operate under such requirements, as they either require substantial amount of buffer space or require the entire data stream~ \cite{douglas_peucker}\cite{Hershberger92speedingup}. Existing online algorithms operate retrospectively on trajectory data or assume favourable trajectory characteristics, resulting in the worst-case complexity of their online performance ranging from $\mathcal{O}(nlogn)$ to $\mathcal{O}(n^2)$~\cite{squishe}. The high complexity of existing methods limits their utility in resource-constrained environments. To address this challenge, we propose the Bounded Quadrant System (BQS), an online algorithm that can run sustainably on resource-constrained devices. Its fast version achieves $\mathcal{O}(n)$ time and $\mathcal{O}(1)$ space complexity while providing guaranteed error bounds. By using a convex hull that is formed by a bounding box and two angular bounds around all points, the algorithm is able to make quick compression decisions without iterating through the buffer and calculating the maximum error in most of the cases. Using empirical GPS traces from the flying fox scenario and from cars, we evaluate the performance of our algorithms and quantify their benefits in improving the efficiency of trajectory compression.

In summary, our contributions are three-fold:
\begin{itemize}
\item Proposal of an efficient online trajectory compression algorithm. The algorithm uses convex hull bounding to compress incoming points into trajectory segments with error guarantees. The fast version of the algorithm achieves constant time and space complexity for each step, or equivalently $\mathcal{O}(n)$ time complexity and $\mathcal{O}(1)$ space complexity for the whole data stream. 
\item Demonstration of the generality and extensibility of the proposed algorithm to support the 3-D case and a different error metric.
\item Comprehensive evaluation of the algorithm using real-life data collected from wildlife tracking and vehicle tracking applications as well as synthetic data.
\end{itemize}

The remainder of this paper is organized as follows. The next section surveys related work. We then motivate the need for a new online trajectory compression algorithm by describing our hardware platform and the data acquisition process. The BQS algorithms are presented subsequently. A discussion on how to generalize the BQS algorithm to a higher dimension is provided. Finally, we evaluate BQS on empirical and synthetic traces, and conclude the paper.

%

%% file: relatedwork.tex
The rapid increase in the number of GPS-enabled devices in recent years has led to an expansion of location-based services and applications and our increased reliance on localization technology. One challenge that location-based applications need to overcome is the amount of data that continuous location tracking can yield. Efficient storage and indexing of these datasets is critical to their success, especially for embedded and handheld devices which provide users with location context in-situ.

Several trajectory compression algorithms that offer significant improvements in terms of data storage have been proposed in the literature. We focus our review on lossy compression algorithms as they provide better trade-offs between high compression ratios and an acceptable error of the compressed trajectory.

Douglas and Peucker were amongst the first to propose an algorithm for reducing the number of points in a digital trajectory~\cite{douglas_peucker}. The Douglas-Peucker algorithm starts with the first and last points of the trajectory and repeatedly adds intermediate points to the compressed trajectory until the maximum spatial error of any point falls bellow a predefined threshold. The algorithm guarantees that the error of the compressed trajectory is within the bounds of the target application, but due to its greedy nature, it achieves relatively low compression ratios. The worst-case runtime of the algorithm is $\mathcal{O}(n^2)$ with $n$ being the number of points in the original trajectory which has been improved by Hershberger et al to $\mathcal{O}(n\log n)$~\cite{Hershberger92speedingup}.

The disadvantage of the Douglas-Peucker algorithm is that it runs off-line and requires the whole trajectory. This is of limited use for modern location-aware applications that require online processing of location data. A generic sliding-window algorithm (conceptually similar to the one summarized in~\cite{opening_window}) is often used to overcome this limitation and works by compressing the original trajectory over a moving buffer. The sliding-window algorithm can run online, but its worst-case runtime is still $\mathcal{O}(nL)$ where $L$ is the maximum buffer size. On the other hand, multiple examples of fast algorithms exist in the literature~\cite{ChenXF12}\cite{Long_2013}. These algorithms, however, do not apply to our scenario as they only run off-line and cannot support location-aware applications in-situ.

SQUISH~\cite{squish} has achieved relatively low runtime, high compression ratios and small trajectory errors. However, its disadvantage was that it could not guarantee trajectory errors to be within an application-specific bound. A follow up work presented SQUISH-E~\cite{squishe} that provides options to both minimize trajectory error given a compression ratio and to maximize compression ratio given an error bound. The worst-case runtime of SQUISH-E algorithms is $\mathcal{O}(n\log\frac{n}{\lambda})$, where $\lambda$ is the desired compression ratio.  While the compression ratio-bound flavor of SQUISH-E can run online, the error-bound version runs offline only.

The disadvantage of these algorithms that repeatedly iterate through all points in the original trajectory is their long runtime. SQUISH-E is approaching linear computational complexity for large compression ratios, however, the resulting compressed trajectory has unbounded error. Our motivation is to develop algorithms that can fit the computational and space constraints of energy-constrained devices and that can be used to track locations of small objects or animals over long time periods with bounded errors. More complex algorithms, such as STTrace~\cite{sttrace} that uses estimation of speed and heading to predict the position of the next location or the MBR algorithm~\cite{liu_MBR} that maintains, divides, and merges bounding rectangle representing the original trajectory fall outside of capabilities of our target hardware platform, which is described in detail in the following section.

In contrast to existing methods, our approach achieves constant time and space complexity for each point by only considering the most recent minimal bounding convex-hulls. We show in the evaluation section that the compression ratios that our approach achieves are superior to those of the related trajectory compression algorithms. Although simplistic approaches such as Dead Reckoning~\cite{Trajcevski06on-linedata}\cite{ Kjaergaard:2009:EER:1555816.1555839} achieve comparable runtime performance, we show that our algorithm significantly outperforms these protocols in compression ratio while guaranteeing an error bound.


%% file: data.tex
\subsection{Motivating Scenario}

We employ the Camazotz mobile sensing platform~\cite{DBLP:conf/ipsn/JurdakSKKCMW13}, which has been specifically designed to meet the stringent constraints for weight and size when being employed for wildlife tracking, in particular for different species of flying foxes, also known as megabats (Pteropus). Animal ethics requires that the total payload weight is smaller than a certain percentage (usually below 5\%) of the animal's body weight, which corresponds to a weight limit of roughly 20-30\,g for flying foxes. Camazotz is a light-weight but feature-rich platform built around the Texas Instrument CC430F5137 system on chip, which integrates a 16-bit microcontroller (32 KBytes ROM, 4 KBytes RAM) and a short-range radio in the 900\,MHz frequency band. We use a rechargeable Lithium-ion battery connected to a solar panel to provide power to the device. Several on-board sensors such as temperature/pressure sensor, accelerometer/magnetometer and a microphone allow for multi-modal activity detection~\cite{DBLP:conf/ipsn/JurdakSKKCMW13} and sensing of the animal's environment. A ublox MAX6 receiver allows to determine the current position using the Global Positioning System (GPS). Sensor readings and tracking data can be stored locally in external flash storage (1 MByte) until the data can be uploaded to a base station deployed at animal congregation areas using the short range radio transceiver. We also employ the same hardware placed on the dashboard of a car to capture mobility traces of vehicles in urban road networks. 

The GPS traces collected by such platforms are often used to model and analyze the mobility and the behavior of the moving object. Hence, it is most important to gather information for the object's major movements. The areas where it often visits, the route it takes to travel between its places of interests, and the time windows in which it usually makes the travels, are the key features we want to extract from the traces. However, the hardware limitations of the platform constrain its capability to capture such information in the long term. Motivated by this application, we propose to use online trajectory compression on such resource-constrained platforms to reduce the data size. The compression will introduce a bounded error and discard some information for small movements, but will capture the interesting moments when major traveling occurs.

\subsection{Existing Solutions}
\subsubsection{Buffered Douglas-Peucker}
Douglas-Peucker (DP) \cite{Heckbert95surveyof} is a well-known algorithm for trajectory
compression. In our scenario, in which the buffer size is largely constrained due to
memory space limit on the platform, we are still able to apply DP on a
smaller buffer of data. We call this algorithm Buffered Douglas-Peucker (BDP) . That is, the incoming points are accumulated in the buffer until it is full, then the partial trajectory defined by
the buffered points is processed by the DP algorithm.  However,
such solution has inferior compression rates mainly due the extra points taken 
when the buffer is repeatedly full, 
preventing a desirable high compression rate from being achieved.

A result of using DP with a fixed-size buffer
is that both the first and last points in the buffer are kept in the
compressed output every time the buffer is full, even when they can actually be discarded safely. In the
worst case scenario where the object is moving in a straight line from
the start to the end, this solution will use $floor(\frac{N}{M})+1$
points, where $N$ is the number
of total points and 
$M$ is the buffer size. In contrast, the optimal solution needs to keep only two points.
 Although the overhead
depends on the shape of the trajectory and the buffer size, generally BDP takes considerably more points than necessary, particularly for small buffer sizes.

\subsubsection{Buffered Greedy Deviation}
Buffered Greedy Deviation (BGD, a variation of the generic sliding-window algorithm) represents another
simplistic approach. In this strategy whenever a point arrives, we
append the point to the end of the buffer, and do a complete calculation of the
error for the trajectory segment defined by the points in the
buffer to the line defined by the start point and the end point. 
If the deviation already exceeds the tolerance, then we keep
the last point in the compressed trajectory, clear the buffer and
start a new segment at the last point. Otherwise the process continues for the next incoming point.

The algorithm is easy to implement and guarantees the error tolerance,
however it too has a major weakness. The compression rate is heavily
dependent on the buffer size, as it faces the same problem as 
BDP. If we increase the buffer size,  because the
time complexity is $\mathcal{O}(n^2)$, the computational complexity would increase
drastically, which is undesirable in our scenario 
because of the energy limitations. Therefore, BGD represents a significant
compromise on the performance as it has to make a direct trade-off between time
complexity and compression rate.

Clearly, a more sophisticated algorithm that can guarantee the bounded error, 
process the point with low time-space complexity, and achieve
high compression rate is
desired. We propose an algorithm called Bounded Quadrant System(BQS) to address this problem. Before delving into the details, we present some
notations, definitions and theorems to help the reader understand
BQS's working mechanism.

%% file: preliminaries.tex
In this section we provide a series of definitions and theorems as the
necessary foundations for further discussion. 

\eat{
\begin{table*}[htp]
\centering
\caption{Symbols and Notations}
\label{tbl:sym}
\begin{tabular}{ | c || p{12cm} | }
\hline Notation & Description \\ 
\hline $v = <lat, lon, t>$ & location at timestamp t \\
\hline $wp = <lat_{min}, lat_{max}, lon_{min}, lon_{max} , w>$ & waypoint area\\
\hline $\tau=\{v_1, ...,v_k\}$ &  trajectory segment $\tau$, composed by a
series of points $v_i$\\ 
\hline $\tau'=\{v_1, v_k\}$ &  compressed trajectory segment $\tau'$\\
\hline  $\mathbb{T}=\{\tau_1,...,\tau_n\}=\{v_1, ...,v_n\}$ & trajectory, each
represents a trip\\
\hline  $\mathbb{T}'=\{v_1, v_j,...v_k\}$ & compressed trajectory, each
adjacent pair $\{v_i,v_j\}$ represent a compressed segment, \newline while together
represents a trip\\
\hline $\varrho = <\beta,\iota>$ & quadrant bounding system\\
\hline $\beta = <min_x, max_x, min_y, max_y>$ & bounding box \\  
\hline $\iota = <l^{left}, l^{right}>$ & left and right angle-bounding
lines \\ 
\hline $\theta_{lb}, \theta{ub}$ &  the smallest and largest angle
from any point in the buffered segment to the start point\\ 
\hline $d^{lb}, d^{ub}$ & the lower bound and
upper bound of the maximum deviation for a quadrant bounding system\\ 
\hline
\end{tabular}
\end{table*}
}

\begin{definition}[Location Point]
A location point $v=<latitude,longitude,$\\$timestamp>$ is a tuple that records the spatio-temporal
information of a location sample.
\end{definition}

\begin{definition}[Segment  and Trajectory]
A trajectory segment is a set of location points that are taken
consecutively in the temporal domain, denoted as $\tau=\{v_1, ...,v_n\}$. A trajectory
is a set of consecutive segments,
denoted as $\mathbb{T}=\{\tau_1,\tau_2,...\}$.
\end{definition}
\eat{
A trajectory represents a trip, which in real-life could be understood
as a journey between two stay points. A stay point is a place at
which the moving object has stayed for a certain time period within a
radius of the place. The stay time and radius are defined according to
the characteristics of the movements for the moving object, such as
the speed or the spatial scale. For instance, car trips tend to have smaller stay
time thresholds and greater radii, as car trips are often
task-related and are of greater spatial scales.
}
Given the definitions of segment and trajectory, we introduce
the concept of compressed trajectory with bounded error: 
\begin{definition}[Deviation]
Given a trajectory segment $\tau=\{v_1, ...,v_n\}$, the 
deviation $\bdd(\tau)$  is defined as the largest distance from any
location $v_i \in \{v_2,...,v_{n-1}\}$ to the line defined by $v_1$ and $v_n$. The
trajectory deviation is defined as the maximal segment deviation from any
of its segments, as $max(\bdd (\tau_i)), \tau_i \in \mathbb{T}$. 
\end{definition}
Deviation is a distance metric to measure the maximum error from the
compressed trajectory segment to the original segment. For
simplicity of the proof and presentation, without loss of generality, 
 we use point-to-line distance in this definition. Note that
 point-to-line-segment distance can be easily used within BQS too.

\begin{definition}[Key Point]
Given a trajectory segment buffer $\tau=\{v_1, ...,v_k\}$ and a
deviation threshold $\epsilon^d$, a new location
point $v_n$, $v_k$ is a key point if $\bdd(\tau) \leq \epsilon^d$ and
$\bdd(\tau \bigcup \{v_n\}) > \epsilon^d$, where $\epsilon^d$ is the
error tolerance.
\end{definition}
In other words, when a new point is sampled, the
immediate previous point is classified as a key point if the new point
results in the maximum error of any point in the segment buffer exceeding $\epsilon^d$.

\begin{definition}[Compressed Trajectory]
Given a trajectory segment $\tau=\{v_1, ...,v_n\}$, its compressed
trajectory segment
is defined by the start and end location $v_1$ and $v_n$, and is denoted
as $\tau'=\{v_1,v_n\}$. The compressed trajectory $\mathbb{T}' =\{v_i, vj, ...,v_k\}$ of
$\mathbb{T}$ is the set of starting and ending locations of all the
segments in $\mathbb{T}$, ordered by the position of its source
segment in the original trajectory.
\end{definition}

\begin{definition}[Error-bounded Trajectory]
An error-bounded trajectory is a compressed trajectory with the 
deviation for any of its compressed segments smaller than or equal to
a given error tolerance
$\epsilon^d$. Formally: 
given a trajectory $\mathbb{T}=\{\tau_1,...,\tau_k\}$, and
its compressed trajectory $\mathbb{T}' =\{v_i, vj, ...,v_k\}$,
$\mathbb{T}'$ is error-bounded by $\epsilon^d$ if $\forall \tau_i'\in \mathbb{T}'$,
$\bdd(\tau_i') \leq \epsilon^d$.
\end{definition}

\begin{figure}[htp]
\vspace{-0.5cm}
\hspace{0.2cm}
\includegraphics[height=4.8cm]{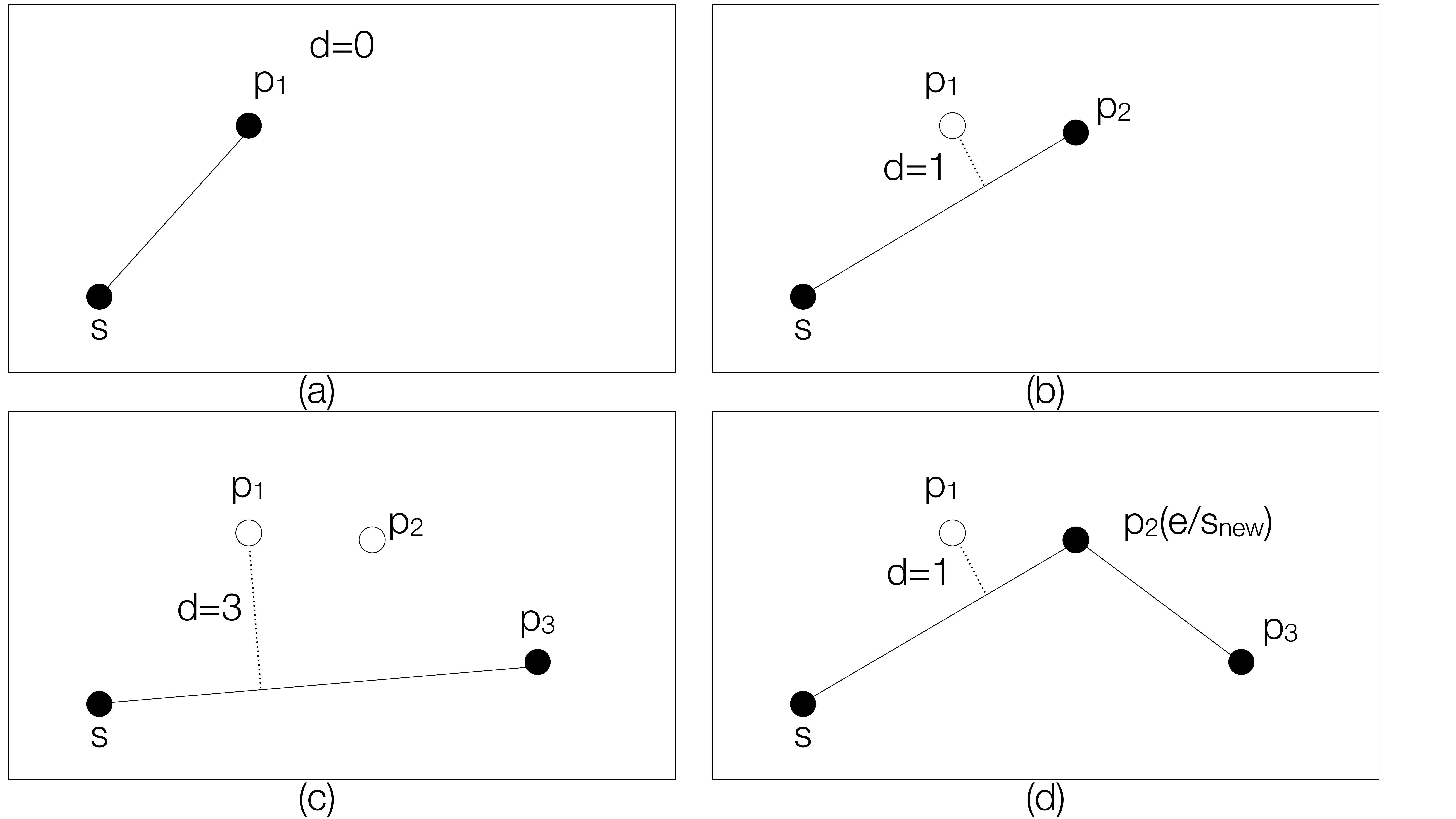}
\vspace{-0.3cm}
\caption{Error-bounded Compression ($\epsilon=2$)}
\vspace{-0.3cm}
\label{fig:cexp}
\end{figure}

Figure \ref{fig:cexp} demonstrates the process of error-bounded
trajectory compression. Assuming that the current trajectory segment
starts from $s$, when adding the first point $p_1$ (Figure
\ref{fig:cexp}(a)), the deviation is $0$. Hence we proceed to the next
incoming point $p_2$ (Figure \ref{fig:cexp}(b)). Here the deviation
is $1$, which lies within the error tolerance, so the current segment can be
safely represented by $\overline{sp_2}$. However after $p_3$ arrives,
the deviation of the segment reaches $3>\epsilon$ because of
$p_1$ (Figure \ref{fig:cexp}(c)). Clearly $p_3$ should not be included in the current
trajectory segment. Instead, the current segment ends at $p_2$ and a
new segment is then started at $p_2$ (Figure \ref{fig:cexp}(d)). The
new segment includes $p_3$ and repeats the above process for new
incoming points until the error tolerance is exceeded again.
Such process guarantees that
any trajectory segment has a smaller deviation than $\epsilon$.

When a trajectory is turned into a compressed trajectory, the temporal
information it carries changes representation too. Instead of
having a timestamp at every location point in the original trajectory,
the compressed trajectory uses the timestamps of the key points as the
anchors to reconstruct the time sequences. Given a trajectory segment
defined by two key points $v_s, v_e$, the reconstructed location at
timestamp $\overline{t}$ ($v_s.t \le \overline{t}\le v_e.t$) is defined as:

\begin{equation}
v_{\overline{t}} = < h_{lat}(\mathrm{P}, v_s, v_e, \overline{t}), h_{lon}(\mathrm{P},
v_s, v_e, \overline{t}), \overline{t} >.
\end{equation}
where function $h$ is an interpolation function that uses a
distribution function $\mathrm{P}$, a start value, an end
value, and a timestamp at which the value should be
interpolated. $\mathrm{P}$ interpolates the location at a timestamp
according to a distribution. As
an example, the
$h$ and $\mathrm{P}$ functions for interpolating the latitude can be defined as:
\begin{eqnarray}
\mathrm{P}(\overline{t}) &=& \frac{\overline{t}-v_s.t}{v_e.t - v_s.t}\\
h_{lat} (\mathrm{P}, v_s, v_e, \overline{t}) &=& v_s.lat +
\mathrm{P}(\overline{t}) \times (v_e.lat - v_s.lat)
\end{eqnarray} 
where $\mathrm{P}$ is set to reconstruct the uniform
distribution. However, in practice this function can be derived online
to fit the distribution of the actual data. For instance, an online algorithm for
fitting Gaussian distribution by dynamically updating the variance and mean can be
implemented with semi-numeric algorithms described in
\cite{Knuth:1997:ACP:270146}, which can be used to derive
$\mathrm{P}$. 
\eat{Again, we want to state that
reconstructing the trajectory flawlessly is not the goal of this
algorithm. The trajectories processed by our algorithm will capture
the dominant movements. The algorithm overall is
designed to capture larger movements while maintaining a reasonable,
guaranteed error tolerance, which is useful for many types of mobility analysis.
}

As we are designing an online algorithm where each point is processed
only once, the problem is turned into answering the following question: does 
the incoming point result in a new compressed trajectory segment or can it be
represented in the previous compressed trajectory segment? To address
this question, we first provide an overview and then the details for the BQS framework.

%% file: overview.tex

The motivation of the algorithm is that we need a trajectory
management infrastructure to store historical trajectory data with
minimal storage space while maintaining the error-bounds and 
capturing the major movements of the mobile object. Hence we need an 
efficient online trajectory compression algorithm, that yields
error-bounded results in low time and space complexity and that minimizes
the number of points taken, i.e. the Bounded Quadrant System (BQS).

\eat{
Th framework aims to address the following:
\begin{enumerate}
\item 
\item We need a waypoint area discovery functionality to facilitate
  advanced tasks like trip time estimation etc. A waypoint is an area
  at where multiple trajectories join paths but part ways afterwards.
\end{enumerate} 
Subsequently, we formulate the sub-problems as follows:
\begin{enumerate}
\item Trajectory Compression: \textit{
Given a limited buffer size and computation power, and a stream
that continues to feed new location points, determine
whether the incoming point should be kept as the part of the current
segment, or the trigger of a new segment, so that the current segment's 
deviation will be bounded by the threshold $\epsilon$.
}

\item Trajectory Management: \textit{
While a trajectory is being processed, and compressed segments are fed
into the framework, the framework should be able to index the segment,
merge similar segments, and further compress aged trajectories to optimize space utilization.
}
\item Waypoint Area Discovery:\textit{
The framework should be able to periodically 
search and retrieve similar historic segments and trajectories so that
small areas which have heavy and diversified historical traffic are discovered as
waypoint areas. 
}
\end{enumerate} 
}
A BQS is a convex hull that bounds a
set of points (an example is shown in Figure \ref{fig:bqs}). A BQS is constructed by the following steps:
\begin{enumerate}
\item For a trajectory segment, we split the space into four
  quadrants, with the origin set at the start point of the current
  segment, and the axes set to the UTM(Universal Transverse Mercator) projected $x$ and $y$ axes.
\item For each quadrant, a bounding box is set for the buffered points
  in that quadrant, if there are any. There can be at most four
  BQS for a trajectory segment.
\item We keep two bounding lines that record
  the smallest and greatest angles between the $x$ axis and the line from the origin to
  any buffered point for each quadrant.
\item We have at most eight significant points in every quadrant systems - four
  vertices on the bounding box, four intersection points from the
  bounding lines intersecting with the bounding box. Some of the points may overlap.
\item Based on the deviations from the significant points to the
  current path line, we have a group of lower bound candidates and
  upper bound candidates for the maximum deviation. From these candidates we derive a pair of lower
  bound and 
upper bound $<d^{lb},~d^{ub}>$, to make compression decisions without the full computation of segment
  deviation in most of the cases. 
\end{enumerate}
Here the lower bound $d^{lb}$
  represents the smallest deviation possible for all the points in the segment
  buffer to the start point, while the upper bound $d^{ub}$ represents the
  largest deviation possible for all the points in the segment
  buffer to the start point. With $d^{lb}$ and $d^{ub}$, we have three
  cases:
\begin{enumerate}
\item If $d^{lb} > \epsilon ^d$, it is unnecessary to perform
  deviation calculation because the deviation is guaranteed to break
  the tolerance, so a new segment needs to be started.
\item If $d^{ub} \leq \epsilon ^d$ , it is unnecessary to perform
  deviation calculation because the deviation is guaranteed to be
  smaller or equal to the tolerance, so the current point will be included in the
  current segment, i.e. no need to start a new segment.
\item If $d^{lb} \leq \epsilon ^d < d^{ub}$, a deviation
  calculation is required to determine whether the actual deviation is breaking
  the tolerance.
\end{enumerate}
Hence a pair of bounds is considered ``tight'' if the
  difference between them is small enough that it leaves minimum
  room for the error tolerance $\epsilon^d$ to be in between them.

%% file: compression.tex
\begin{figure}[htp]
\vspace{-0.3cm}
\hspace{1.2cm}
\includegraphics[height=3.8cm]{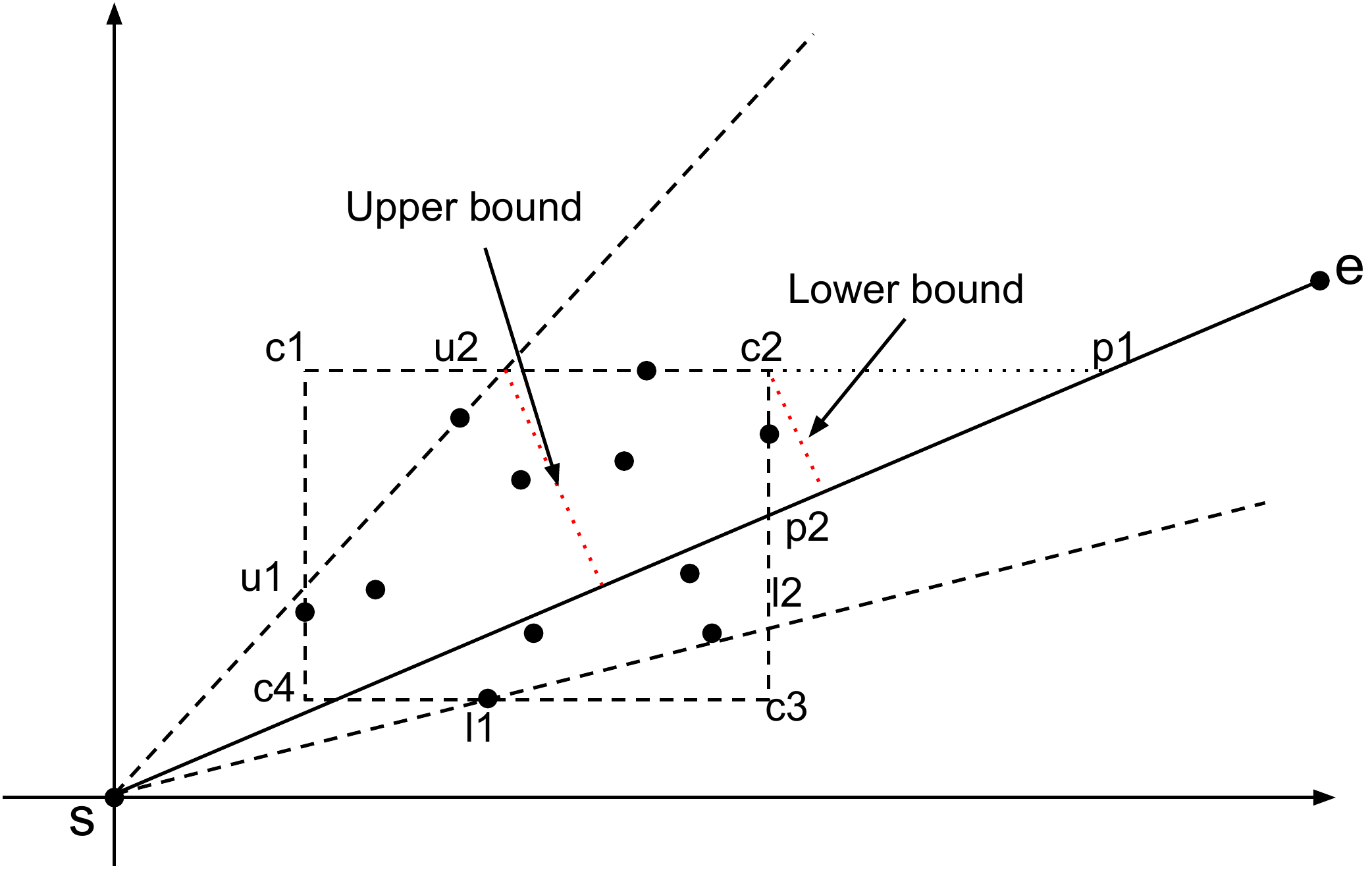}
\caption{An Example of the BQS}
\label{fig:bqs}
\end{figure}

\vspace{-1cm}
\subsection{The Bounded Quadrant System}


An illustration of a BQS is provided in Figure \ref{fig:bqs}. Here we have
a BQS in the first quadrant. Note that it shows only one BQS but in
reality there could be at most four BQS for each segment, one for each
quadrant. In the
figure, $s$ is the start point of the current segment, which is also
used as the origin of the BQS. The solid
black dots are the buffered points for the current segment. The bounding box
$\overline{c_1c_2c_3c_4c_1}$ is the minimum bounding rectangle for all
the points in the first quadrant. The bounding lines
$\overline{su_{2}}$ and $\overline{sl_{2}}$ record the greatest and
    smallest angles that any line from the origin to the points can
    have (w.r.t. the $x$ axis), respectively.

The intuition of the BQS structure is to exploit the favorable properties
of the convex hull formed by the
significant points from the bounding box
and bounding lines for the buffered points (excluding the start point). That is, with the
polygons formed by the bounding box and the bounding lines, we
can derive a pair of tight lower bound and upper
bound on the deviation from the points in the buffer to the line
defined by the start and end points denoted as $l_{s,e}$. 
With such bounds, most of the
deviation calculations on the buffered points are avoided. Instead we can
determine the compression decisions by only assessing a few significant
vertices on the bounding polygon. The splitting of the space into four
quadrants is necessary as it guarantees a few useful properties to
form the error bounds. More details are provided in Section \ref{sec:app}.

To understand how the BQS works, we use the example
with a start point $s$, a few points in the buffer, and the last incoming
point $e$ as in Figure \ref{fig:bqs}. The goal is to determine whether
the deviation will be greater than the
threshold if we include 
the last point in the current segment. First we present two fundamental theorems:

\begin{theorem}
\label{thm:-1}
Assume that a point $p$ satisfies $d(p, s)\leq \epsilon$, where
s is the start point, then
\begin{eqnarray}
d^{max}(p, l_{s,e}) \leq \epsilon
\end{eqnarray}
regardless of the location of the end point $e$.
\end{theorem}
Proof: Trivial.
This theorem enables the quick decision on an incoming point even
without assessing the bounding boxes or lines. Such a point is directly ``included''
in the current segment, and the BQS structure will not be affected.
\begin{theorem}
\label{thm:0}
Assume that the buffered points $\{p_i\}$ are
bounded by a rectangle in the spatial domain, with the vertices $c_1, c_2,
c_3, c_4$, if we denote the line defined by the start
point $s$ and the end point $e$ as $l_{s,e}$, then we always have:
\begin{eqnarray}
d^{max}(p_i, l_{s,e}) \geq min\{ d(c_i, l_{s,e}) \} = d^{lb}\\
d^{max}(p_i, l_{s,e}) \leq max\{ d(c_i, l_{s,e}) \} =d^{ub}
\end{eqnarray}
\end{theorem}
Proof: The proof of this theorem is fairly straightforward if we
regard the polygon $\overline{c_1c_2c_3c_4c_1}$ as a convex hull. Nevertheless we
give an alternative proof in details to help readers understand the concept.  
The bounding box
dictates that on each of its edges there must be at least one point.
If for any edge $\overline{c_jc_k}$ of the bounding box we denote a
buffered point on it as $p^e$, then we have $ min\{ d(c_j, l_{s,e}), d(c_k, l_{s,e})\} \leq d(
p^e, l_{s,e} ) \leq max\{ d(c_j, l_{s,e}), d(c_k, l_{s,e})\}
$. Consolidating the bounds on all of the edges, we have the proof of
the theorem.

Theorems \ref{thm:-1} and \ref{thm:0} show how some of the points can
be safely discarded and how the basic lower bound and upper bound properties are
derived. However, Theorem \ref{thm:0}
only provides a pair of loose bounds that can hardly avoid any deviation
computation. To obtain tighter and useful bounds, we need to introduce
a few
advanced theorems. Throughout the theorem definitions we use the following notations:

\begin{itemize}
\item Corner Distances: We use $d^{corner} = \{ d(c_i, l_{s,e})\},
  ~ i\in\{1,2,3,4\}$ to denote the distances from each vertex of
  the bounding box to the current path line.
\item Near-far Corner Distances: We use $d^{corner-nf} = \{ d(c_n, l_{s,e}), d(c_f, l_{s,e})\},
  $ to denote the distances from the nearest vertex $c_n$ and the
  farthest vertex $c_f$ of the bounding box (near and far in terms of
  the distance to the origin)  to the current path
  line. 
  The nearest and farthest corner points are determined by the
  quadrant the BQS is in. For example, in Figure \ref{fig:bqs} $c_n=c_4$ and $c_f=c_2$. 
\item Intersection Distances: We use $d^{intersection} = \{ d(p,
  l_{s,e})\},  ~ p \in\{l_1, l_2, u_1, u_2\}$ to
  denote the distances from each 
intersection point to the current path line, where $l_i$ are the
  intersection points of the lower angle bounding line and the bounding box,
  and $u_i$ are the
  intersection points of the upper angle bounding line and the bounding box.
\end{itemize}

Some advanced bounds are defined as follows:

\begin{theorem}
\label{thm:1}
Given a BQS, if the line $l_{s,e}$ is in the quadrant, and
$l_{s,e}$ is in between the two bounding lines ($\theta_{lb} \leq
\theta_{s,e} \leq \theta_{ub}$), then we have the following bounds on
the segment deviation: 
\begin{eqnarray}
d^{max}(p, l_{s,e}) &\geq& d^{lb} = \\ \nonumber &max& \begin{cases} ~ & min\{ l_1, l_{s,e}),
  d(l_2,l_{s,e}) \}  \\ ~ & min\{ d(u_1, l_{s,e}),
  d(u_2,l_{s,e}) \} \\ ~ & max\{d^{corner-nf}\}  \end{cases}\\
\label{eqn:ub} d^{max}(p, l_{s,e}) &\leq&  d^{ub} = max\{  d^{intersection} \}
\end{eqnarray}

\end{theorem}
A line $l$ is ``in'' the quadrant $Q$ if the angle $\theta^l$ between $l$ and the
$x$ axis satisfies $ \theta^Q_{start}
\leq \theta^l < \theta^Q_{end}$ or $ \theta^Q_{start}
\leq \theta^l + \pi< \theta^Q_{end}$ or $ \theta^Q_{start}
\leq \theta^l - \pi< \theta^Q_{end}$, where $\theta^Q_{start}$ and
$\theta^Q_{start}$ are the angle range of the quadrant where the BQS
resides. Note that this definition is distance metric-specific. Since
we use point-to-line distance, a line is automatically ``in'' exactly
two quadrants.
In future references we assume
$\theta^l$ satisfies $ \theta^Q_{start}
\leq \theta^l < \theta^Q_{end}$ if it is ``in''
the quadrant.

\begin{theorem}
\label{thm:2}
Given a BQS, if the line $l_{s,e}$ is in the quadrant, and
$l_{s,e}$ is outside the two bounding lines ($\theta_{ub} <
\theta_{s,e} ~or~ \theta_{lb} > \theta_{s,e}$), we have the same bounds on
the segment deviation as in Theorem \ref{thm:1}.
\end{theorem}

\eat{
\begin{theorem}
\label{thm:3}
Given a QBS,, then we have the same bounds on
the segment deviation as in Theorem \ref{thm:1}.
\end{theorem}

Theorems \ref{thm:1},\ref{thm:2},\ref{thm:3} are used in
difference cases when the path and the bounding boxes as well as the
bounding lines are showing different layouts. Though the three
theorems provide the same lower bounds and upper bounds, 
we present them in separate cases because the layout affects how the
bounds are generated and how they can be proven. But in general, if
the path line is in the quadrant, it should be applicable with one of
the three theorems.
}
If the path line is not in the same quadrant with the BQS, we use
Theorem \ref{thm:4} to derive the bounds:
\begin{theorem}
\label{thm:4}
Given a BQS, if the line $l_{s,e}$ is not in the quadrant, 
the bounds of the segment deviation are defined as: 
\begin{eqnarray}
d^{max}(p, l_{s,e}) &\geq& d^{lb} = \\ \nonumber &max& \begin{cases} ~ & min\{ d(l_1, l_{s,e}),
  d(l_2,l_{s,e}) \}  \\ ~ & min\{ d(u_1, l_{s,e}),
  d(l_2,l_{s,e}) \} \\ ~ & 3^{rd}-largest(d^{corner})  \end{cases}\\
d(p, l_{s,e}) &\leq&  max\{  d^{corner} \} = d^{ub}
\end{eqnarray}
\end{theorem}
Proof of the theorems is provided in the appendix.

\eat{A summary of the bounds and criteria is provided in the
following table:
\begin{table}[htp]
\caption{Bounds and conditions}
\label{tbl:bnd}
\begin{tabular}{ | c |c|c|c | }
\hline \shortstack{In\\quadrant} & \shortstack{Intersects\\box} &
\shortstack{In bounding\\lines} & \shortstack{Applicable\\ Theorem} \\ 
\hline $\surd$ & $\surd$ & $\surd$ & \ref{thm:1} \\
\hline $\surd$ & $\surd$ & $\times$ &\ref{thm:2} \\
\hline $\surd$ & $\times$ & $\times$ & \ref{thm:3} \\
\hline $\times$ & $\times$ & $\times$ & \ref{thm:4} \\
\hline
\end{tabular}
\end{table}
Note that the conditions here are progressive. A line must be in the
quadrant to be able to intersect with in the bounding box, then it has
be intersecting with the bounding box to be in the bounding
lines. Given such relations, the table here covers every possibility
that can occur given a line and a QBS, which guarantees in every case
there will be applicable lower bound and upper bound to help avoid an
actual full calculation of segment deviation.}

\subsection{The BQS Algorithm}

The BQS algorithm is formally described in Algorithm \ref{alg:1}:

\begin{algorithm}[thp]
\caption{The BQS Algorithm}
\textbf{Input:} Start point $s$, incoming new point $e$, buffered
points $\mathcal{B}$, deviation threshold $\epsilon^d$\\
\textbf{Algorithm:}

{\begin{algorithmic}[1]

  \IF{$d(s,e) \leq \epsilon^d $}
  \STATE $Decision$: $e\rightarrow \mathcal{B}$
  \ELSE
  
  \STATE Assume $e$ is the end point, compute the lower bound \\
  $d^{lb}_i$ and upper bound $d^{ub}_i$ of the maximal deviations \\
  for the points in each quadrant using Theorems \ref{thm:1},\\
  \ref{thm:2}, \ref{thm:4}
  \STATE Aggregate all the lower bounds and upper bounds in each BQS by computing.
  $d^{lb} = max\{ d^{lb}_i \}$ and   $d^{ub} = max\{ d^{ub}_i \}$
  
  \IF{$d^{ub} \leq \epsilon^d $}
  \STATE $Decision$: $e\rightarrow \mathcal{B}$
  
  \ELSIF{$d^{lb} > \epsilon^d $}
  \STATE $Decision$: Current segment stops and new \\segment starts at the previous point before $e$

  \ELSIF{$d^{lb} \leq \epsilon^d < d^{ub}$}
  \STATE $d = ComputeDeviation(\mathcal{B}, l_{s,p})$
  \STATE $Decision$: made according to $d$
  \ENDIF
  \ENDIF
  \STATE Determine which quadrant $e$ lies in and update the
  \\bounding structure for the corresponding BQS

\end{algorithmic}
}
\label{alg:1}
\end{algorithm}
The algorithm starts by checking if there is a trivial decision: by
Theorem \ref{thm:-1}, if the incoming point $e$ lays within the range of
$\epsilon$ of the start point, no matter where the final end point is,
the point will not result in a greater deviation than
$\epsilon$ (Lines 1-3). After $e$ passes this test, it means $e$ may result in a
greater deviation from the buffered points. So we assume that $e$ is
the new end point, and assume the current segment is presented by
$l_{s,e}$. Now for each quadrant we have a BQS, maintaining their
respective bounding boxes and bounding lines. For each BQS, we have a
few (8 at most) significant points, identified as the four corner
points of the bounding box and the four intersection points between
the bounding lines and the bounding box. According to the theorems we
defined, we can aggregate four sets of lower bounds and upper bounds
for each quadrant, and then a global lower bound and a global upper
bound for all the quadrants (Lines 4-5). According to the global lower
bound and upper bound, we can quickly make a compression decision. If
the upper bound is smaller than $\epsilon$, it means no buffered point
will have a deviation greater than or equal to $\epsilon$, so the
current point $e$ is immediately included to the current trajectory
segment and the current segment goes on (Lines 6-7). On the contrary, 
if the lower bound is greater than
$\epsilon$, we are guaranteed that at least one buffered point will
break the error tolerance, so we save the current segment (without
$e$) and start a new segment at $e$ (Lines 8-9). Otherwise if the
tolerance happens to be in between the lower bound and the upper
bound, an actual deviation computation is required, and decision will
be made according to the result (Lines 10-13). Finally, if the
current segment goes on, we put $e$ into its corresponding quadrant
and update the BQS structure in that quadrant (Line 14).

\begin{figure}[htp]
\vspace{-0.3cm}
\hspace{0.8cm}
\includegraphics[height=5cm]{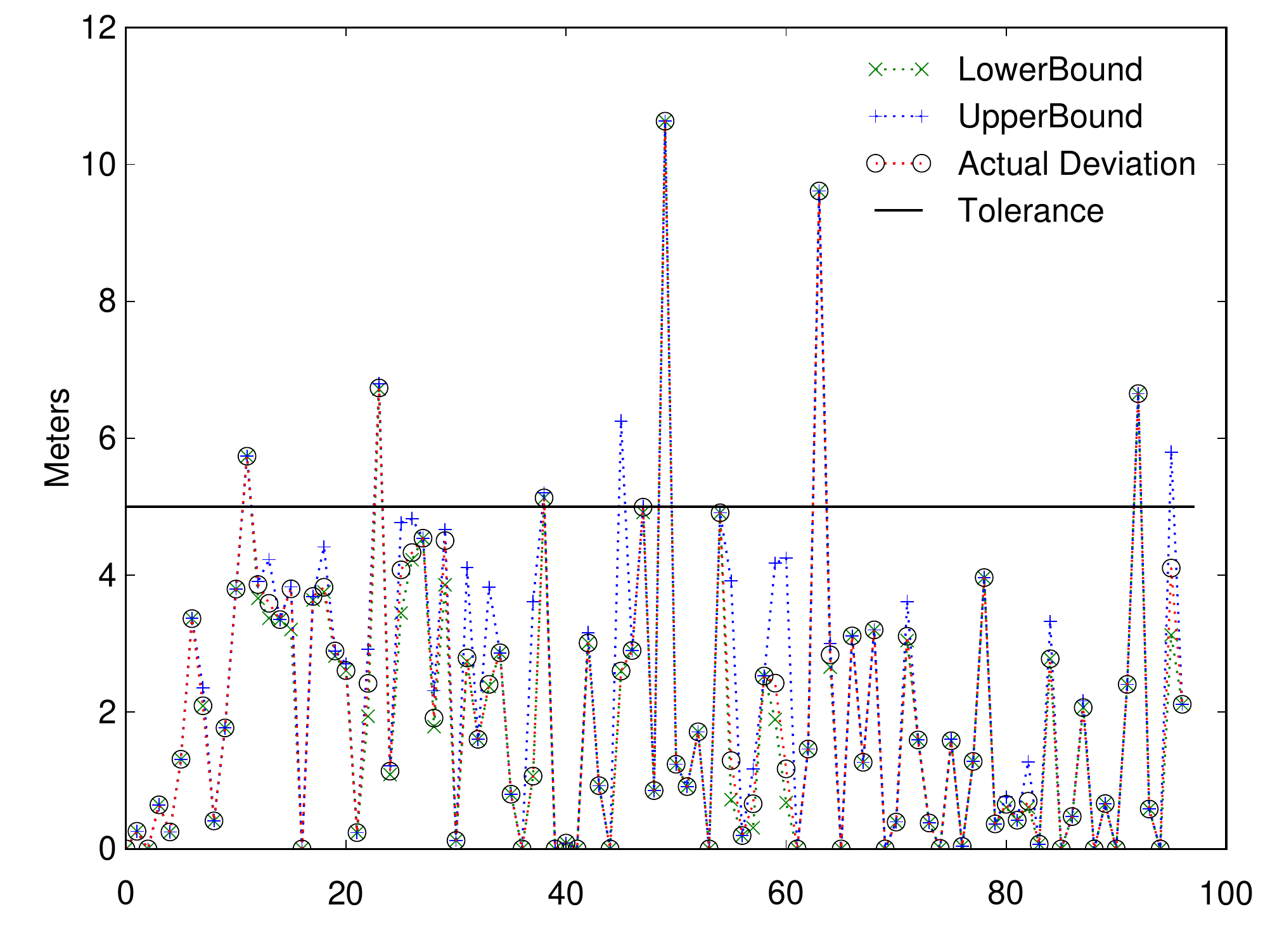}
\caption{Bounds v.s. Actual Deviation}
\label{fig:bds}
\end{figure}
\vspace{-0.7cm}

Figure \ref{fig:bds} demonstrates the lower and upper
bounds of some randomly-chosen location points from the real-life
flying fox dataset, with
$\epsilon^d$ set to $5~m$. The $x$ axis shows the indices of the points,
while the solid horizontal line indicates the error tolerance. It is evident that
in most cases the bounds are very tight and that in more than $90\%$ of the
occasions we can determine if a point is a key point by using only the
bounds and avoid actual deviation calculations.

\subsection{Data-centric Rotation}
A technique called data-centric rotation is used to further tighten 
the bounds. When a new segment is started, instead of constructing and 
updating the BQS immediately after the arrival of new points, we allow 
a tiny buffer to store the first few 
points (e.g. 5) that are not in the range of $\epsilon$ within the start point
(meaning that these points will actually affect the bounding box). 
With the buffer, we compute the centroid of the buffered
points, and rotate the current $x$ axis to the line from the start
point to the centroid. By applying this rotation, we enforce that the 
points are split into two BQS. This improves the
tightness of the bounds because the bounding convex-hulls are
generally tighter with less spread. Once the rotation angle is
identified, each new point for the same segment is temporally 
rotated by the same angle when estimating their distances to the line
$l_{s,e}$.

\begin{figure}[htp]
\vspace{-0.4cm}
\hspace{1.5cm}
\includegraphics[width=4.7cm]{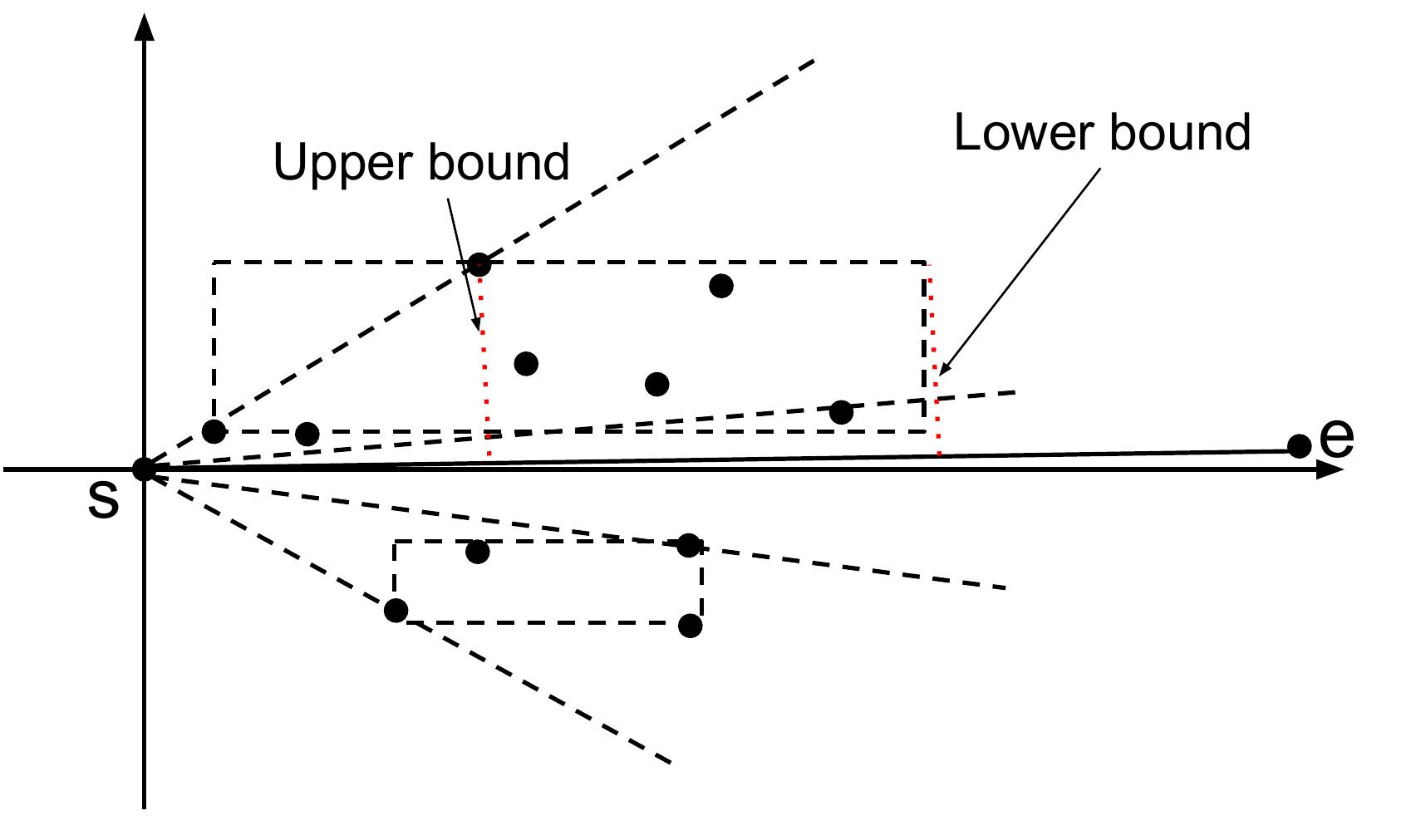}
\caption{Data-centric Rotation}
\label{fig:rtt}
\end{figure}
\vspace{-0.2cm}
Figure \ref{fig:rtt} shows the effect of this rotation. The
points are the same as in Figure \ref{fig:bqs}, but in this figure
they are rotated to ``center'' the data points to the $x$ axis. After
the rotation the points are split into two BQS and it's visually
evident that the gap between the lower bound and the upper
bound becomes smaller. As in reality the likelihood is 
substantial for a moving object to travel in a major direction despite
slight heading changes, this step improves the BQS's pruning power 
significantly. The procedure is applied on Algorithm \ref{alg:1}.

%% file: compression_fast.tex
With the pruning power introduced by the deviation bounds, Algorithm
\ref{alg:1} achieves excellent performance in terms of time
complexity.
Its expected time complexity is $ \alpha \times
n \times c_1 +
(1-\alpha) \times n \times m \times c_2$, where $\alpha$ is the
pruning power, $m$ is the maximum buffer size, and $c_1,c_2$ are two
constants denoting the cost of the processing of each point.
\eat{
\begin{algorithm}[thp]
\caption{Faster BQS Algorithm}
\textbf{Input:} Start point $s$, incoming new point $e$, previous
point $e_p$, deviation threshold $\epsilon^d$\\
\textbf{Algorithm:}

{\begin{algorithmic}[1]

  \STATE Identical with Algorithm \ref{alg:1} before Line 12 except
  line\\ 5 is no longer required
  \IF{$d^{lb} \leq \epsilon^d \leq d^{ub}$}
  \STATE $Decision$ = Current segment stops and new \\segment starts at the previous point $e_p$
  \ENDIF
  \RETURN $Decision$
\end{algorithmic}
}
\label{alg:2}
\end{algorithm}}
Empirical study in the next section shows that $\alpha$ is
generally greater than $0.9$, meaning the time complexity is
approaching $\mathcal{O}(n)$ for the whole data stream. However, the
theoretical worst
case time complexity is still $\mathcal{O}(n^2)$. Moreover, because we 
keep a buffer for the points in the current segment for potential
deviation calculation, the worst-case space complexity
is $\mathcal{O}(n)$. To further reduce the complexity,  
we propose a more efficient algorithm that still utilizes
the bounds but is able to completely avoid any full deviation
calculation and any use of buffer, making the time and space
 costs of processing a point constant.

The algorithm is nearly identical to Algorithm \ref{alg:1}.
The major difference
is that whenever the case $d^{lb} \leq \epsilon^d < d^{ub}$ occurs
(Line 10),
an aggressive approach is taken. No deviation calculation is
performed, instead we take the point and start a new trajectory
segment to avoid any computation and to eliminate the necessity of
maintaining a buffer for the points in the current segment. So 
Lines 11-12 in Algorithm \ref{alg:1} are changed into making the ``stop and
restart'' decision (as in Line 9) directly without any full calculation in Line
12. Also, the maintenance of the buffer is not needed any more.

The Fast BQS (FBQS) algorithm takes slightly more points than Algorithm
\ref{alg:1} in the compression, reducing the compression rate by a small margin. However,
the simplification on the time and space complexity is
significant. The fast BQS algorithm achieves constant complexity in both time and
space for processing a point. Equivalently, time and space
complexity are $\mathcal{O}(n)$ and $\mathcal{O}(1)$ for
the whole data stream. 

The time complexity is only introduced
by assessing and updating a few key variables, i.e. bounding lines, intersection points and
corner points. We can now arrive at the compression decision by keeping only the
significant BQS points of the number $c\leq 32$ ( 4 corner points and
4 intersection points at most for each quadrant) for the
entire algorithm.

When the buffer size is unconstrained, the three algorithms 
Buffered Douglas-Peucker (BDP),
Buffered Greedy Deviation (BGD) and Fast BQS (FBQS)
have the following worst-case time and space complexity:
\begin{table}[htp]
\centering
\caption{Worst-case Complexity}
\label{tbl:cmp}
\begin{tabular}{ | c |c|c| }
\hline  & \textbf{Time} & \textbf{Space} \\ 
\hline \textbf{FBQS} & $\mathcal{O}(n)$ & $\mathcal{O}(1)$\\
\hline \textbf{BDP} & $\mathcal{O}(n^2)$& $\mathcal{O}(n)$\\
\hline \textbf{BGD} & $\mathcal{O}(n^2)$& $\mathcal{O}(n)$\\
\hline
\end{tabular}
\end{table}
\vspace{-0.5cm}
\eat{
Note that when the compressed trajectories are inserted to the spatial
index, their bounding boxes are loosened to cover the error-bound, so
that whenever a search is performed on the index, results will not be
missed because of the deviation introduced in the compression.
}

\subsection{Maintenance Procedures}
In addition to the compression itself, the BQS framework employs two
techniques to further reduce the storage space required for the
historical data, namely error-bounded merging and error-bounded
ageing. Due to space limitations, we only give a brief description of the
techniques. More details will be incorporated in an extended version
of this work. 

Merging is a procedure in which the newly compressed
segment is used as a query to search
similar historical segments in the trajectory database. If any
existing compressed segment could represent the same path with a minor error,
the new segment is considered duplicate information and is merged
into the existing one. 

Ageing is based on the intuition that newer and 
older trajectories should not
bear the same significance in the historical trajectory database.
More recent trips represent the moving object's recent travel
patterns better and should be regarded of greater interest.
Hence the ageing procedure re-runs the compression algorithm on the existing
trajectories that are already compressed, but with a greater error
tolerance, so that the compression rate will be further improved.

The procedures are aligned with our goal, namely
capturing the spatio-temporal characteristics such as the time windows,
destinations and routes for the major movements of the moving
object. By making a more aggressive trade-off between accuracy and
coverage, it further extends the frameworks capability to capture
mobility patterns over long periods.

%% file: generalization.tex
\subsection{Generalization}
\label{sec:gen}
The BQS algorithm's advantage is not only in its competitive
complexity and compression rate, but also in its extensibility to
derive other variations for different application requirements. In
this section we briefly introduce a 3-D variant of the BQS algorithm to
demonstrate its extensibility for more complex application
requirements such as 3-D location tracking and time-sensitive tracking.

In some scenarios such as indoor tracking or aero vehicle tracking, 
the tracking process takes places in a
3-D geographic space \cite{citeulike:3939699}. At any timestamp, 
the moving object has not only
the latitude and longitude coordinates, but also the altitude
component. The simplification error is then determined by the maximum
deviation from the original trajectory in this 3-D space defined by $<latitude, longitude, altitude>$.

In some location tracking applications it is desirable to know where
the moving object is at a certain time. Therefore, time-sensitive error such as
in \cite{Cao:2006:SDR:1147679.1147681} becomes a useful error
metric in such scenarios. Under such settings, 
instead of deriving the deviation from the
original trajectory solely in the $<latitude, longitude>$ plane, 
the error is also attributed by the time axis 
in a three dimensional space $<latitude, longitude, timestamp>$. 

\begin{figure}[htp]
\vspace{-0.4cm}
\hspace{1.2cm}
\includegraphics[height=5cm]{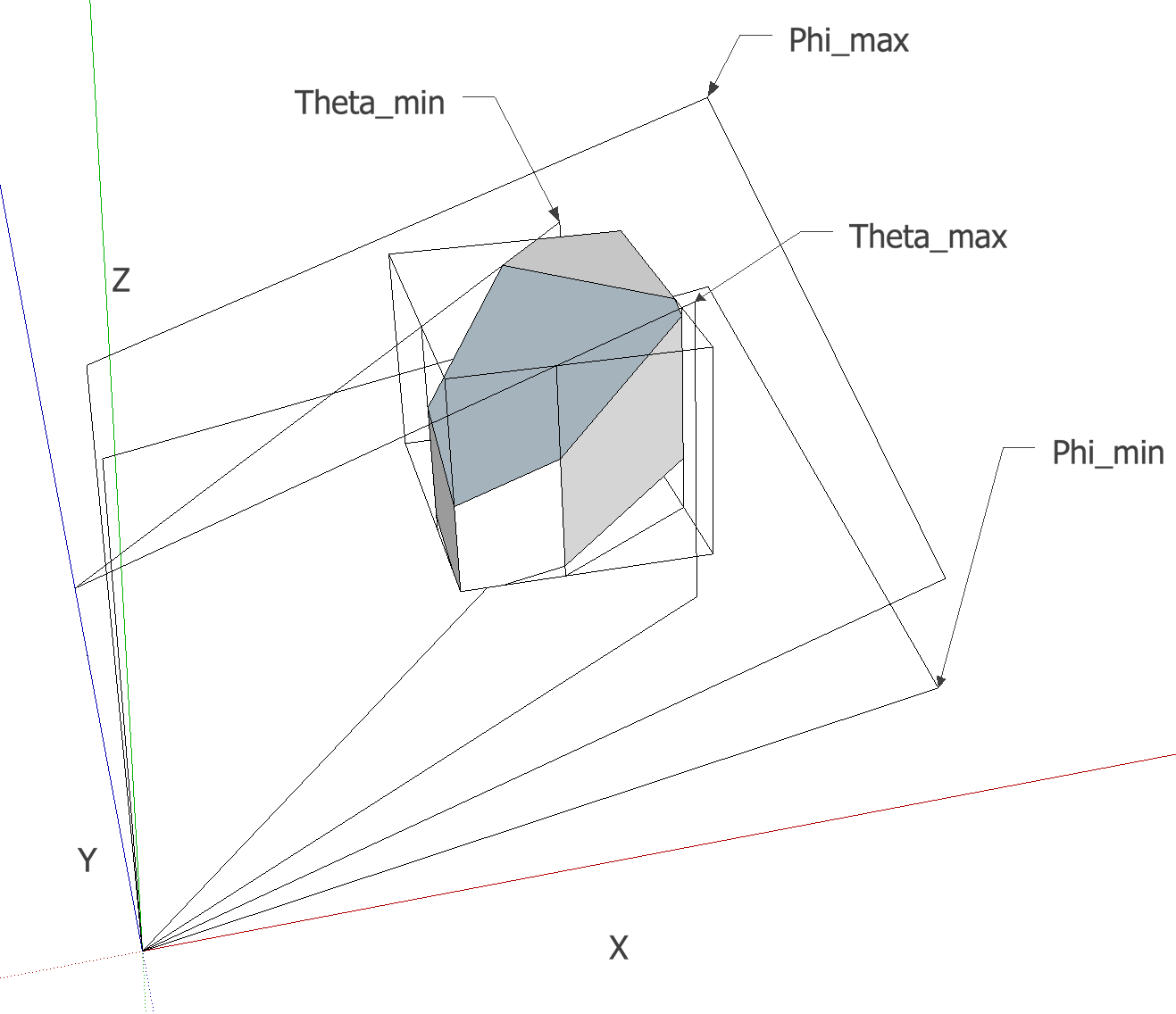}
\caption{The 3-D version of BQS illustrated}
\label{fig:bqs3d}
\end{figure}

Both applications require a trajectory compression algorithm that can handle
3-D data. Now we show that extending the BQS algorithm to the 3-D
space is a straightforward process. Let us revisit the example we
discussed in Figure \ref{fig:bqs}, where we demonstrated
the concept of bounding box and bounding line in the 2-D case. For
the 3-D case, instead of using bounding boxes and bounding lines, 
we use bounding right rectangular prisms and bounding
  planes to bound the location points. An illustration
is given in Figure \ref{fig:bqs3d}.

\vspace{-0.3cm}

In Figure \ref{fig:bqs3d}, we use a Cartesian coordinate system
to represent the 3-D space. Any location point is defined by a
coordinate tuple $<x,y,z>$, where the $z$ axis can either represent
the timestamp or the altitude. The deviation metric is also extended
to measure the maximum distance from original 3-D points to a
3-D line. 

There are eight quadrants in total. We use the first quadrant 
($x>0 \land y>0 \land z>0 $) as a demonstration in the figure. 
A bounding prism is used to bound the location points in each
quadrant, as a direct extension to the 2-D case. Then, in each
quadrant, we also establish two pairs of bounding planes to maintain
the minimum and maximum angles to their respective reference planes. 
We have a pair of ``vertical'' bounding planes $\Theta_{min},\Theta_{max}$, which both
are orthogonal to the $XY$ plane and contain the $z$ axis. They
represent the minimum and maximum angles formed by any plane that contains a
location points to the $YZ$ plane. Similarly, we have the ``inclined''
bounding planes $\Phi_{min}$ and $\Phi_{max}$. They represent the minimum and
maximum angles between the planes in $\{\Phi_i\}$ to the $XY$
plane. Any $\Phi_i$ is determined by three points, namely two anchor
points and one data point, and all $\Phi_i$ in a quadrant share the
same anchor points. The anchor points are determined by the quadrant,
as $(sign(x)\times 1,-sign(y) \times 1,0),(-sign(x) \times 1,sign(y) \times 1,0)$. For this quadrant, the two anchor points
are $(1,-1,0)$ and $(-1,1,0)$.

\eat{
which form two inverted cones
starting from the origin. Every point on a cone yields the same angle
between the line from the origin to it and the line from the origin to
its projected point on the $XY$ plane. However, surfaces for inverted
cones are more expensive for the intersection calculations than
planes are. In practice, we hence use two planes two approximate the two
surfaces, namely the $\Phi_{min}$ and $\Phi_{max}$ bounding planes in
Figure \ref{fig:bqs3d}. They are derived as shown in Figure
\ref{fig:bqs3d}. First we calculate the intersecting curves between
the two cones and the bounding prism, then we choose the approximation
planes separately for the minimum and maximum angles. Figure
\ref{fig:approx} shows the approximation process, from an orthogonal
perspective from a point on the z-axis to the bounding prism. The two dotted
arcs are the intersections of the cones to the prism, and the two
dashed lines are the approximation plane for the two cones. For the maximum
case ($\Phi_{max}$), we use the two intersection points between the
cone and the edges of the first plane}

The bounding prism is hence ``cut'' into a 
3-D polyhedron that is also a convex hull (shadowed part in Figure \ref{fig:bqs3d}), and the
vertices that form the hull are the significant points we use to
quickly derive error bounds. There are
mature software libraries to enable the efficient calculation of the bounding
polyhedron, such as GEOS \footnote{\url{http://trac.osgeo.org/geos/}}
or CGAL \footnote{\url{http://www.cgal.org/}}. 
In practice, to further improve
the efficiency of the computation of the convex hull, we only consider
the intersection points between the bounding planes and the
bounding prism, while intersections between bounding planes
are not considered. This will slightly increase the volume of the
bounding polyhedron but the computational cost will decrease
considerably and become stable (independent of the
spatial relations among bounding planes). Finally, 
we obtain a 3-D convex hull formed by at
most 17 points (at most 4 intersection points for each bounding plane, 
plus the farthest vertex to the origin of the prism). 
Using similar techniques to Theorems
\ref{thm:1}, \ref{thm:2} and \ref{thm:4}, new pruning 
rules for the 3-D BQS can then be derived based on the significant
points. The computation involved for identifying the significant
points is more complex than that in the 2-D
case, but overall the algorithm still has constant time and space
complexity in its fast version setup.

\eat{We also note that the
two bounding planes may cut the same part of the prism, resulting
complex intersections at in the figure. In practice this effect is
ignored for computational efficiency. Only the intersections between the
edges of the prism and the four bounding planes are considered as
significant points, leaving at most 16 significant 
points to assess. }
It is worth noting that besides the extended version in the 3-D
space, the BQS algorithm can also be used with different distance
metrics, such as the point-to-line-segment distance. 
In cases where line-to-segment distance is used, Theorems
\ref{thm:1} and \ref{thm:2} can be used with minor
modification by changing Equation \ref{eqn:ub} to:
\begin{equation}
d^{max}(p, \overline{s,e}) \leq d^{ub} = max\{  d^{intersection}, d^{corner-nf} \}
\end{equation}
while Theorem \ref{thm:4} still holds. The definition
of the ``in quadrant'' property is slightly changed accordingly.

%% file: experiments.tex
\section{Experiments}
In this section we evaluate the performance of the
proposed BQS compression framework.

\vspace{-0.2cm}
\subsection{Dataset}
We use three datasets, namely the flying fox (bat) dataset, the vehicle dataset, and
the synthetic dataset. The two real-life datasets comprise of $138,798$ GPS samples, collected by 6 Camazotz
nodes (five on bats and one on a vehicle). The total travel distances for
the bat dataset and vehicle dataset are $7,206~km$ and $1,187~km$
respectively. The tracking periods lasted six months for the bat dataset and
two weeks for the vehicle dataset.

\eat{
\begin{figure}[htp]
\hspace{-0.6cm}
\begin{tabular}{c}
 \subfigure[Bat Tracking Data]{
\includegraphics[height=2cm]{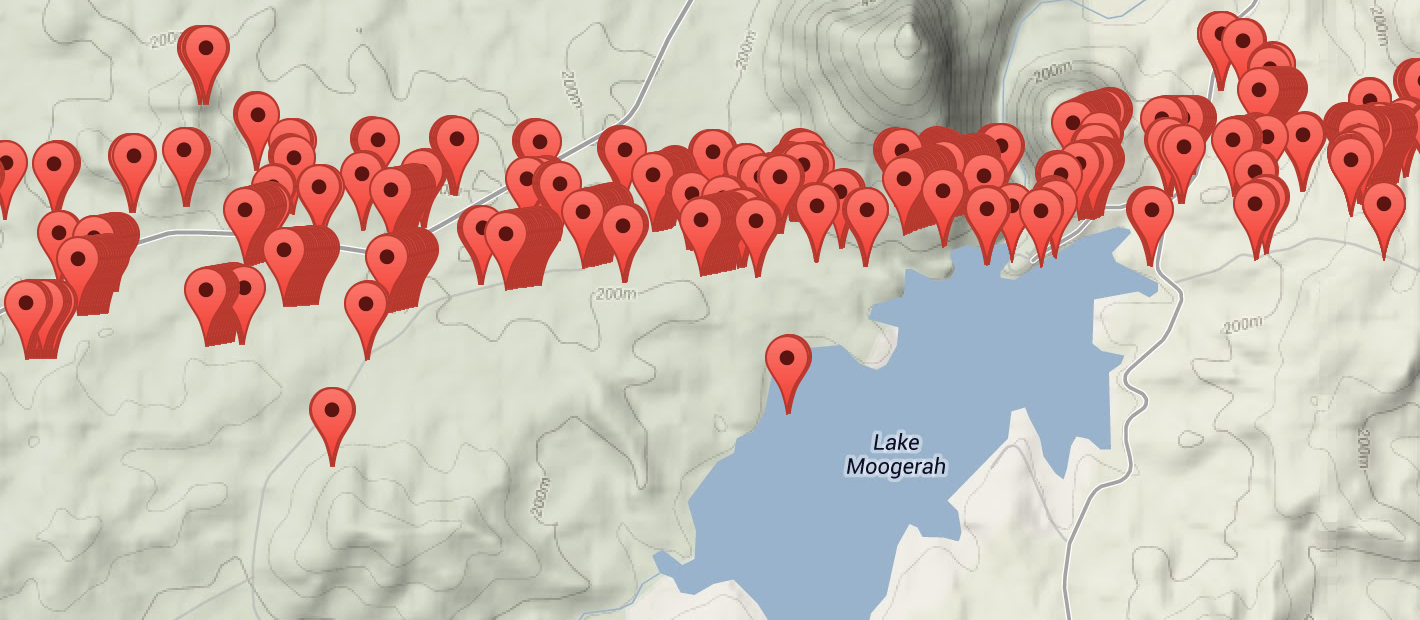}
}
\subfigure[Vehicle Tracking Data]{
\includegraphics[height=2cm]{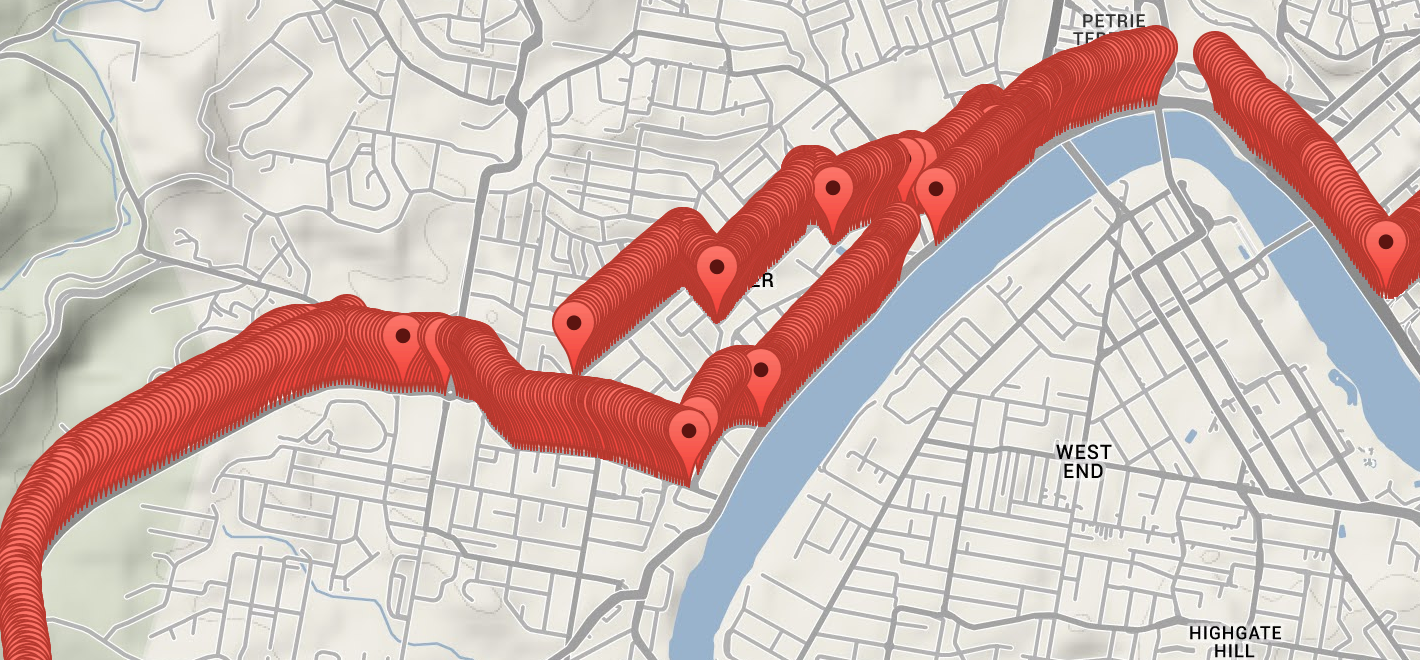}
}
\end{tabular}
\caption{Sample Traces}
\label{fig:sp}
\end{figure}
}
Note that there are couple of differences between the two
datasets. The vehicle dataset shows larger scales in terms of travel distances well as moving speed. For instance, the length of a car
trip varies from a few kilometers to $1,000~km$ while a
trip for flying-foxes are usually around $10~km$. The car can travel
constantly at $100~km/h$ on a highway or $60~km/h$ on common roads,
while the common and maximum continuous flying speeds
for a flying-fox are approximately $35~km/h$ and $50~km/h$. In regard
to these differences, the two datasets are evaluated with different
ranges of error tolerance. With a much greater spatial scale of the
movements, the error tolerance used for the vehicle dataset is generally greater.

The vehicle dataset also shows more consistency in the heading
angles due to the
physical constraints of the road networks. On the contrary,
the bats's movements are unconstrained in the 3-D space, so their
turns tend to be more arbitrary. \eat{Figure \ref{fig:sp} shows 
segments of the actual trajectories from both datasets.}
We argue that by performing extensive experiments on both
datasets, the robustness of our framework to the shape of data is demonstrated.

The synthetic dataset is generated by a statistical model that anchors
patterns from real-life data, and it is used specifically for the
comparison between Dead Reckoning~\cite{Trajcevski06on-linedata} and
Fast BQS (FBQS). This is because continuous
high-frequency samples with speed readings are required to implement
DR in an error-bounded setting, while such data is lacking in the
real-life datasets. The model uses an event-based correlated 
random walk model to simulate the movement of the object. In the
simulation, waiting events and moving events are executed
alternately. The object stays at its previous location during a
waiting event, and it moves in a randomly selected speed and turning
angle for a randomly selected time. Note that the speed
follows the empirical distribution of speed, the turning angle 
is drawn from the von Mise distribution \cite{Risken}, while the move time is 
exponentially distributed, corresponding to the Poisson process. The
trajectories are bounded by a rectangular area of $10~km\times 10~km$,
and the speed and turning angle follow approximately the distributions
of the bat data. A total of $30,000$ points are generated by the
model. 

\subsection{Experimental Settings}
The evaluation is done on a desktop computer, however the extremely 
low space and time complexity of the FBQS algorithms makes it 
plausible to implement the algorithms on the platform aforementioned
 in Section \ref{sec:mot} (32 KBytes ROM, 4 KBytes RAM). Particularly, if we
 look into the FBQS algorithm, we only need tiny memory space to store
 at most 32 points besides the program image itself (4 corner points 
and 4 intersection points for each quadrant).

Two performance indicators,
namely compression rate and pruning power are tested on the real-life datasets.
We define compression rate as $\frac{N^{compressed}}{N^{original}}$ where $N^{compressed}$ is
the number of points after compression, and $N^{original}$ is the number of
points in the original trajectory. Pruning power is defined as $1
-\frac{N^{computed}}{N^{total}}$, where $N^{computed}$ and $N^{total}$
are the number of full deviation calculations and the number of total
points respectively. 

For compression rate, we perform comprehensive comparative study to show
BQS's superiority over the other three methods, namely
Buffered-DouglasPeucker (BDP), Buffered-Greedy (BGD) and Douglas
Peucker (DP). DR is compared against FBQS on the synthetic
dataset. For buffer-dependent algorithms, we set the buffer size to be
32 data points, the same as the memory space needed by the FBQS
algorithm to hold the significant points. 

To intuitively demonstrate the advantage of FBQS in compression rate, 
we then provide performance comparison showing the number
of points taken by the FBQS algorithm and the DR algorithm 
on the synthetic dataset. We also show the estimated
operational time of tracking devices based on such compression rate. Finally,
we study comparatively the actual run time efficiency of FBQS.

For all datasets, we combine all the data points into a single data
stream and use it to feed the algorithms. Then we calculate the
pruning power, compression rate and number of points used.

\begin{figure}[htp]
\vspace{-0.5cm}
\hspace{-0.7cm}
\begin{tabular}{c}
 \subfigure[Bat Tracking Data]{ \label{fig:pp1}
\includegraphics[height=3.5cm]{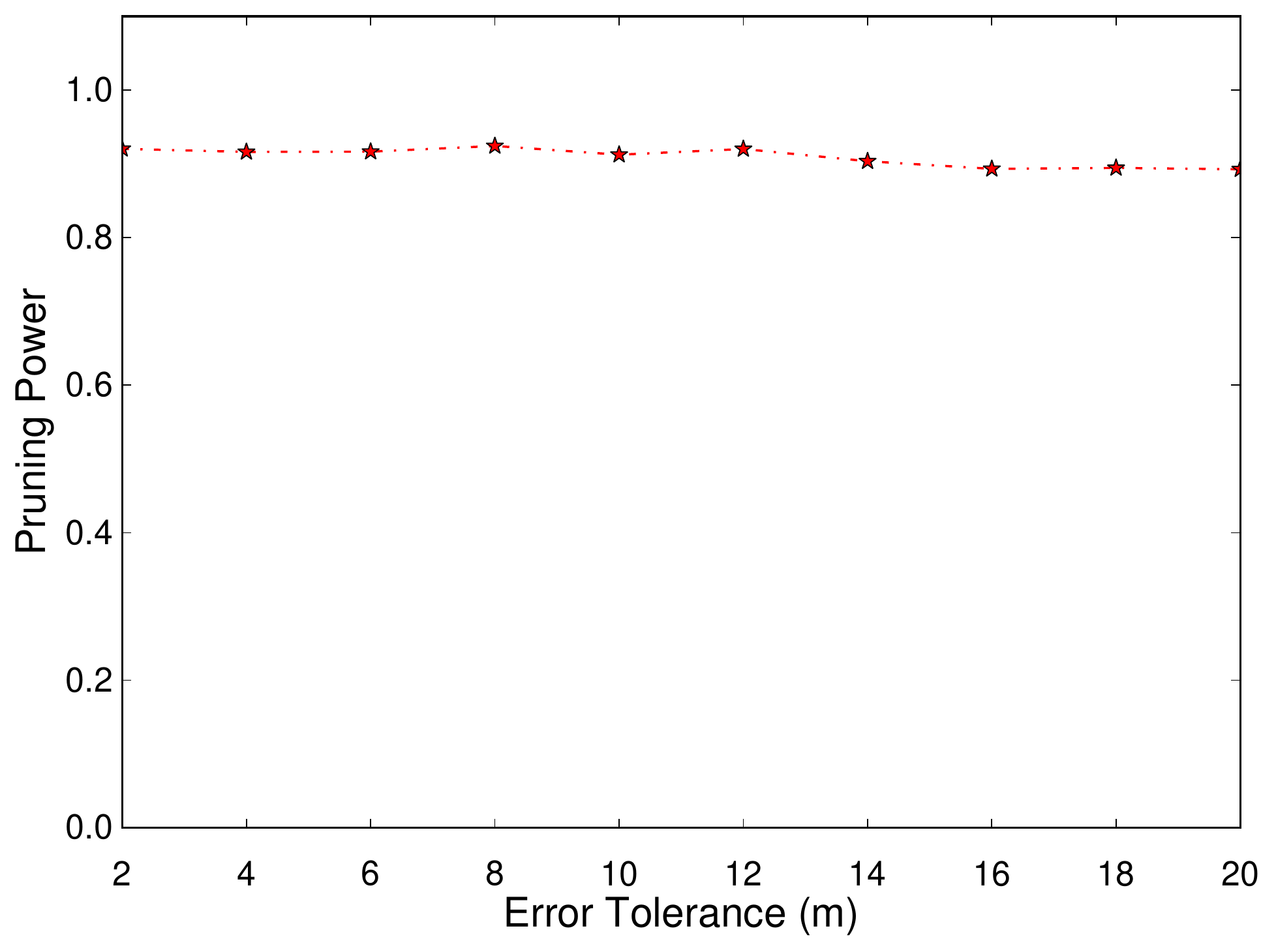}
}
\hspace{-0.5cm}
\subfigure[Vehicle Tracking Data]{ \label{fig:pp2}
\includegraphics[height=3.5cm]{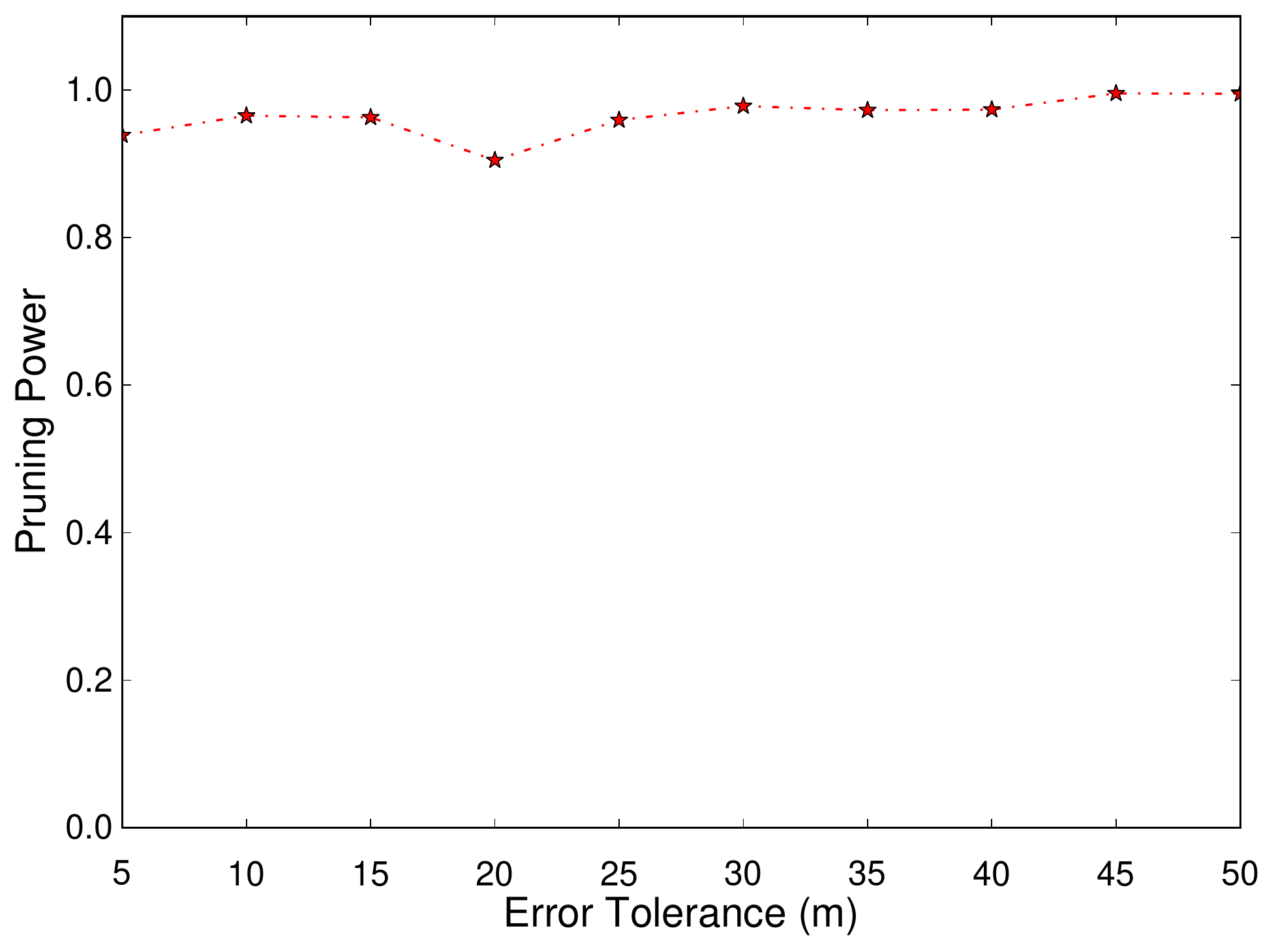}
}
\end{tabular}
\caption{Pruning Power of the BQS Algorithm (The higher the better)}
\vspace{-0.8cm}
\end{figure}

\subsection{Compression Algorithm}

\subsubsection{Pruning Power}

Pruning power determines how efficient BQS and FBQS are. In FBQS, if
the relation between the bounds and the deviation tolerance is
deterministic, FBQS generates a lossless result as
BQS. If it is uncertain, then FBQS will
take a point regardless of the actual
deviation. The pruning power reflects how often the relation is
deterministic, and it indicates how many extra points could be taken by
the approximate algorithm. With high pruning power, 
the overhead of FBQS will be small. Here we investigate
the pruning power of BQS in this subsection. 

Figures \ref{fig:pp1} and \ref{fig:pp2} show the pruning power
achieved by BQS on both datasets. The sensitivity of the algorithm to the error tolerance or to the
shape of the trajectories appears low, 
as the pruning power generally stays above $90\%$ for most of the
tolerance values on both datasets. This means approximately only $10\%$ more
points will be taken in the Fast BQS algorithm compared to the
original BQS algorithm. The running values in Figure \ref{fig:bds} also support
 this observation.

Generally, BQS shows higher pruning power
on the car dataset than on the bat dataset. The higher pruning power on the vehicle data is a
result of the physical constraints of the road networks,
preventing abrupt turning and deviations, and making the trajectories
smooth. Naturally the pruning power will be higher as a result of the higher
regularity in the data's spatio-temporal characteristics.

\begin{figure*}[htp]
\vspace{-0.6cm}
\hspace{0.8cm}
\begin{tabular}{c}
\subfigure[Bat Tracking Data]{ \label{fig:cr1}
\includegraphics[height=5cm]{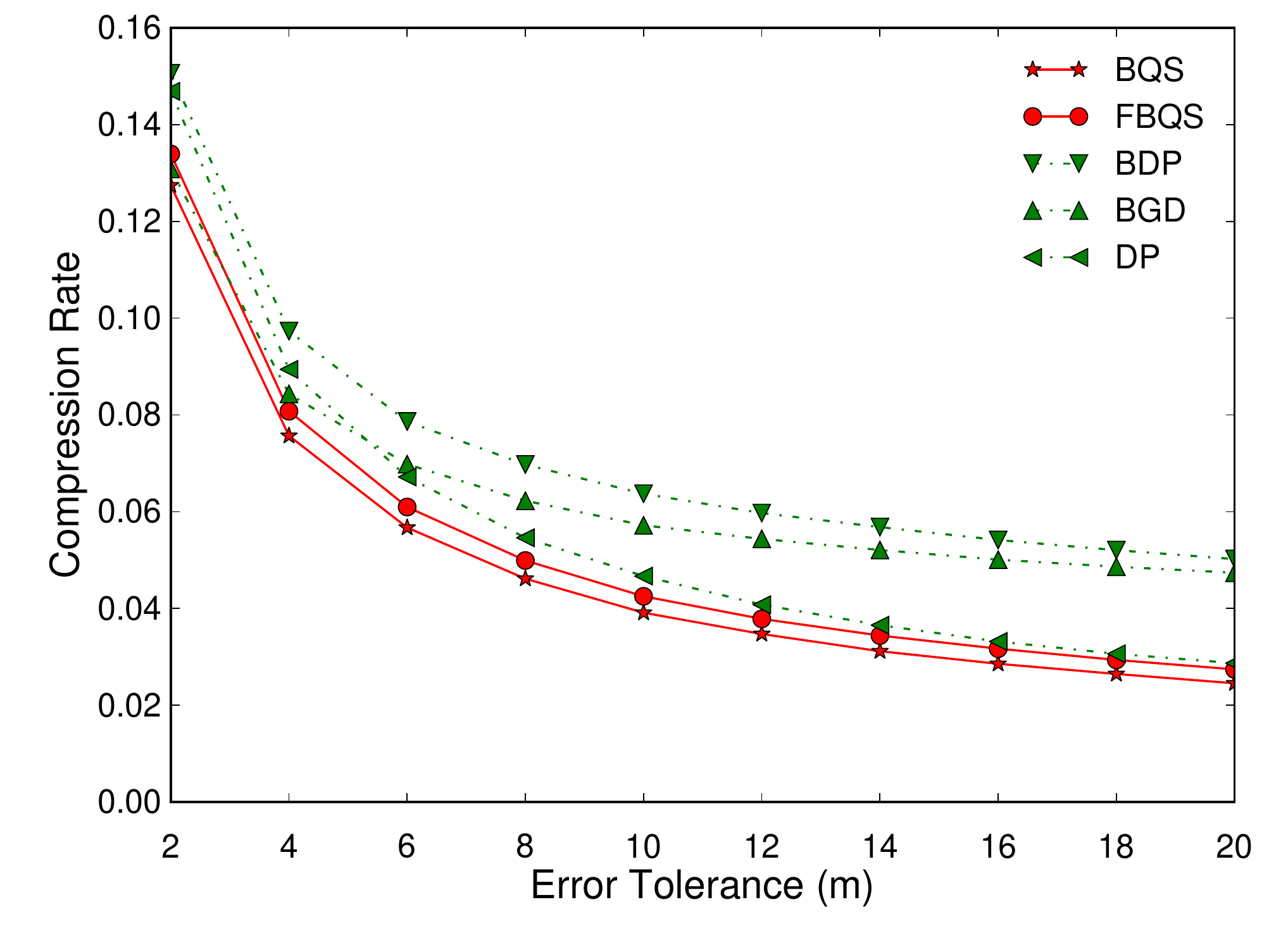}
}\hspace{1.5cm}
\subfigure[Vehicle Tracking Data]{ \label{fig:cr2}
\includegraphics[height=5cm]{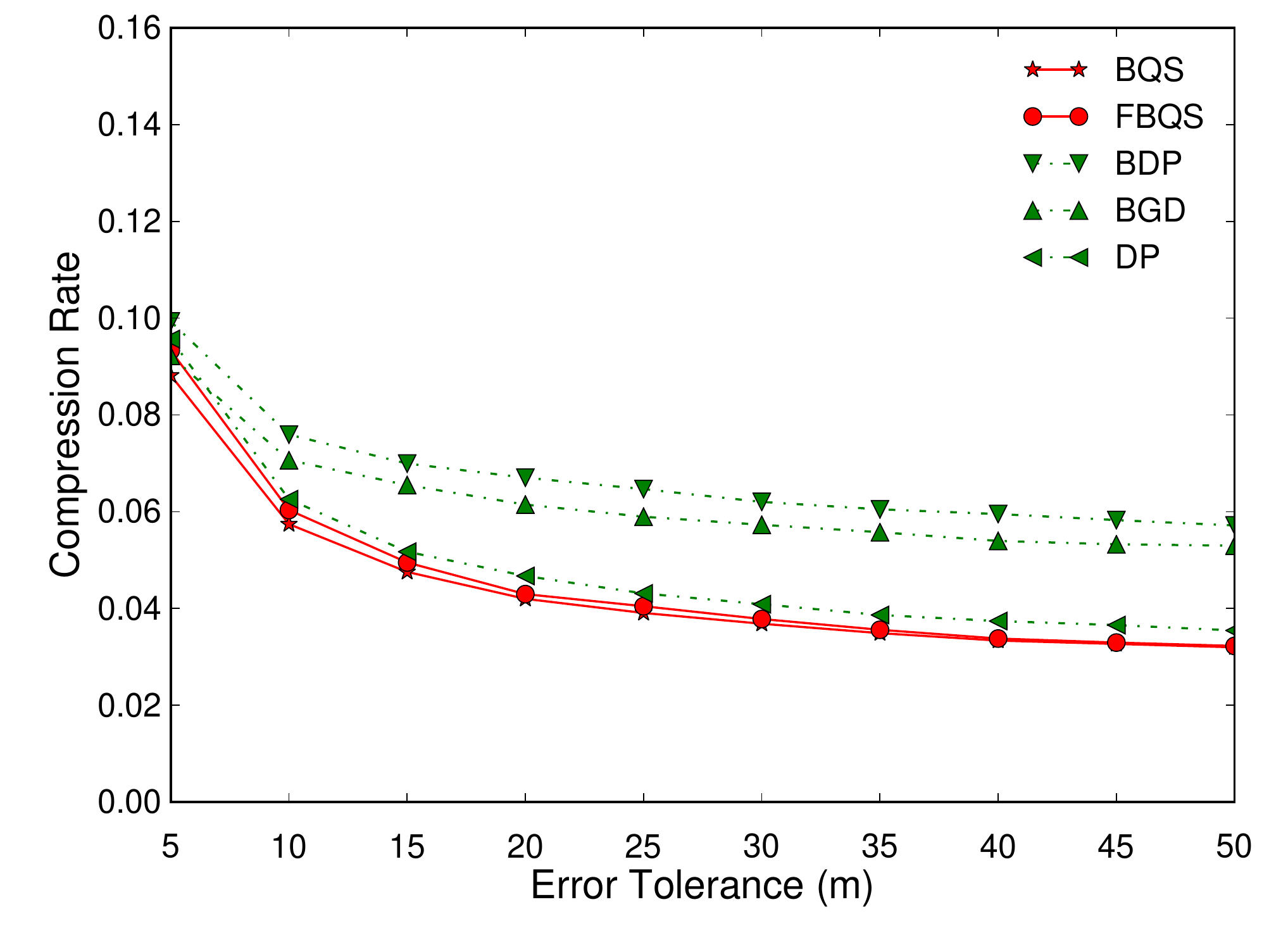}
}
\end{tabular}
\caption{Comparison of Compression Rate on Real-life Datasets (The lower the better)}
\vspace{-0.8cm}
\end{figure*}

\subsubsection{Compression Rate on Real-life Data}

Compression rate is a key performance indicator for trajectory compression
algorithms. Here we conduct tests on the two real-life datasets. We
compare the performance of five algorithms, namely BQS, FBQS,
BDP, BGD and DP. All of the algorithms give error-bounded results.
The former four are online algorithms, and the last one is offline. 
The compression rates are illustrated in Figures \ref{fig:cr1} and \ref{fig:cr2}.

Evidently, BQS achieves the highest compression rate among the five
algorithms, while BDP and BGD constantly use approximately $30\%$ to $50\%$ more
points than BQS does. FBQS's compression rates swing between BQS's
and DP's, showing the second best overal performance. 
BDP has the worst performance overall as it inherits the 
weaknesses from both DP and window-based approaches. BGD's
performance is generally in between DP and BDP, but it still suffers
from the excessive points taken when the buffer is full.

Comparing the two figures, it is worth noting that all the algorithms
perform generally better on the bat data. Take the results at
$10~m$ error tolerance from both figures for example, on the bat data
the best and worst compression rates reach $3.9\%$ and $6.3\%$
respectively, while on the vehicle data the corresponding figures are
$5.4\%$ and $7.7\%$. This may seem to contradict the results of the
pruning power. However, it is in fact reasonable because bats perform 
stays as well as small movement around certain locations, making those
points easily
discardable. Hence the room for compression is larger for the bat
tracking data given the same error tolerance.

On the bat data, with $10~m$ error tolerance, BQS and FBQS achieve
compression rates of $3.9\%$ and $4.1\%$ respectively. DP, as an
offline algorithm that runs in $\mathcal{O}(nlogn)$ time, yields a
 worse compression rate than FBQS at $4.6\%$. Despite having poorer
 worst-case complexities, BDP and BGD also obtain worse compression rates 
than BQS and FBQS do at $6.3\%$ and $5.8\%$ respectively.
At this tolerance, the offline DP
algorithm uses approximately $20\%$ and $10\%$ more points than the
online BQS
and FBQS do, respectively. Furthermore, for online algorithms with $20~m$ tolerance,
FBQS (2.7\%) improves BDP (5.1\%) and BGD (4.9\%) by 47\% and 45\% respectively.

The results on the vehicle data show very similar trends
of the algorithms' compression rate curves. Interestingly, with this 
dataset, because the pruning power is in most of the cases
around and above $0.95$ as demonstrated in \ref{fig:pp2}, the
compression rate of FBQS is remarkably close to BQS'. For instance, at
$20~m$, $30~m$ and upwards, the difference between the two is smaller than $1\%$.
This observation supports our aforementioned argument
that the bounds of the original BQS are so effective that the
 number of extra points taken by FBQS is insignificant.

\eat{
As discussed in Section \MakeUppercase{Trajectory
  Compression with Bounded Quadrant System}, it is expected that BQS has
better compression rate than the two buffered online algorithms. Here
we explain why it also outperforms the offline
DP algorithm. DP's compression rate is heavily
affected by the dividing step. When it decides to split the trajectory
into two parts at the farthest point to the line defined by the start and
end points, there is a great likelihood that the split point is
in the middle of a smooth segment which could be represented by a
single compressed segment. However by making the split, DP takes more
points than BQS as BQS always tracks continuous smooth segments to
minimize the number of points taken. 
}
\vspace{-0.4cm}
\begin{figure}[htp]
\hspace{-0.7cm}
\begin{tabular}{c}
\subfigure[Synthetic Dataset (unit:m)]{ \label{fig:syn}
\includegraphics[height=4cm]{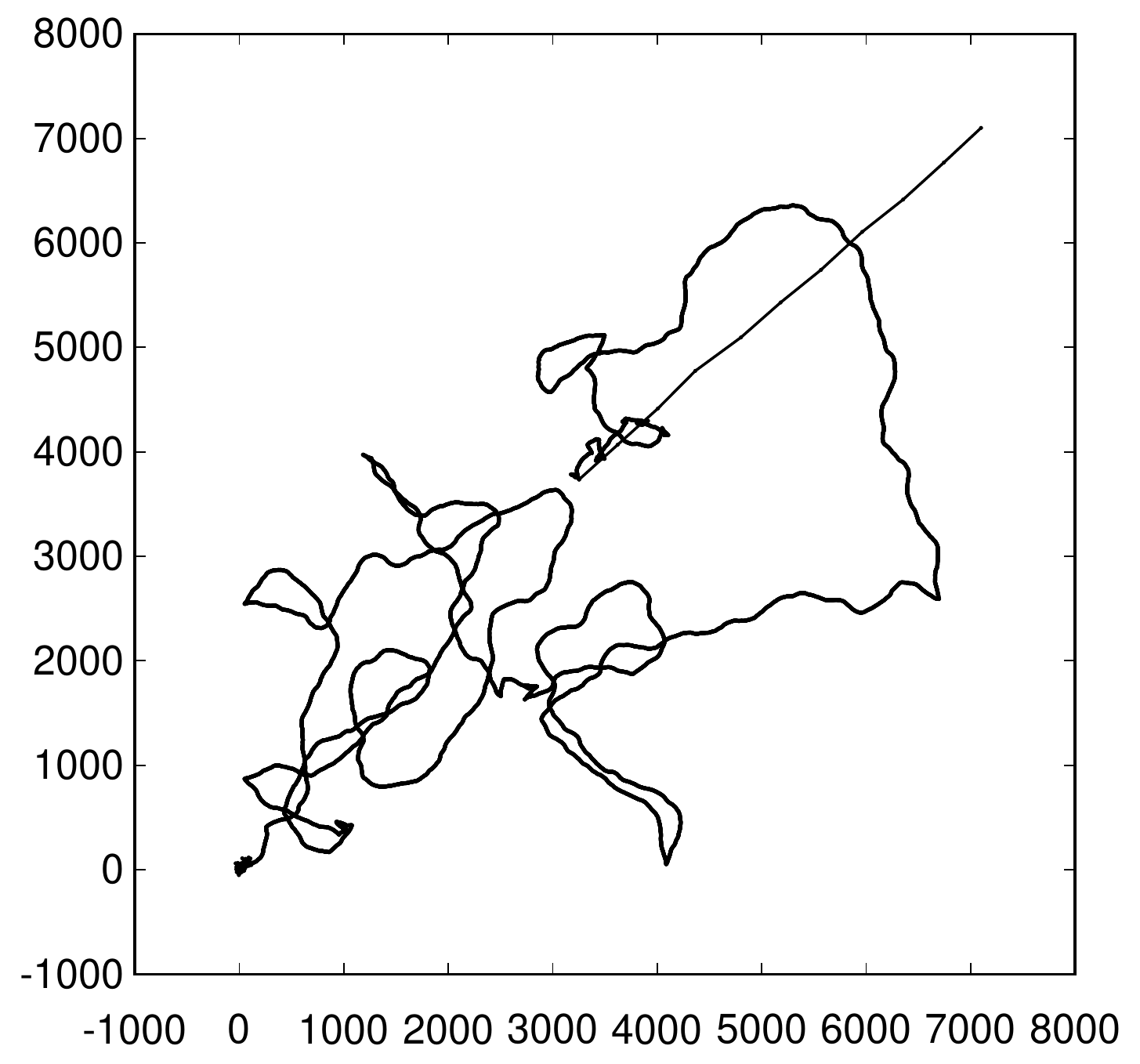}
}
\hspace{-0.7cm}
\subfigure[No.Points Used on Synthetic Data]{ \label{fig:nps}
\includegraphics[height=4cm]{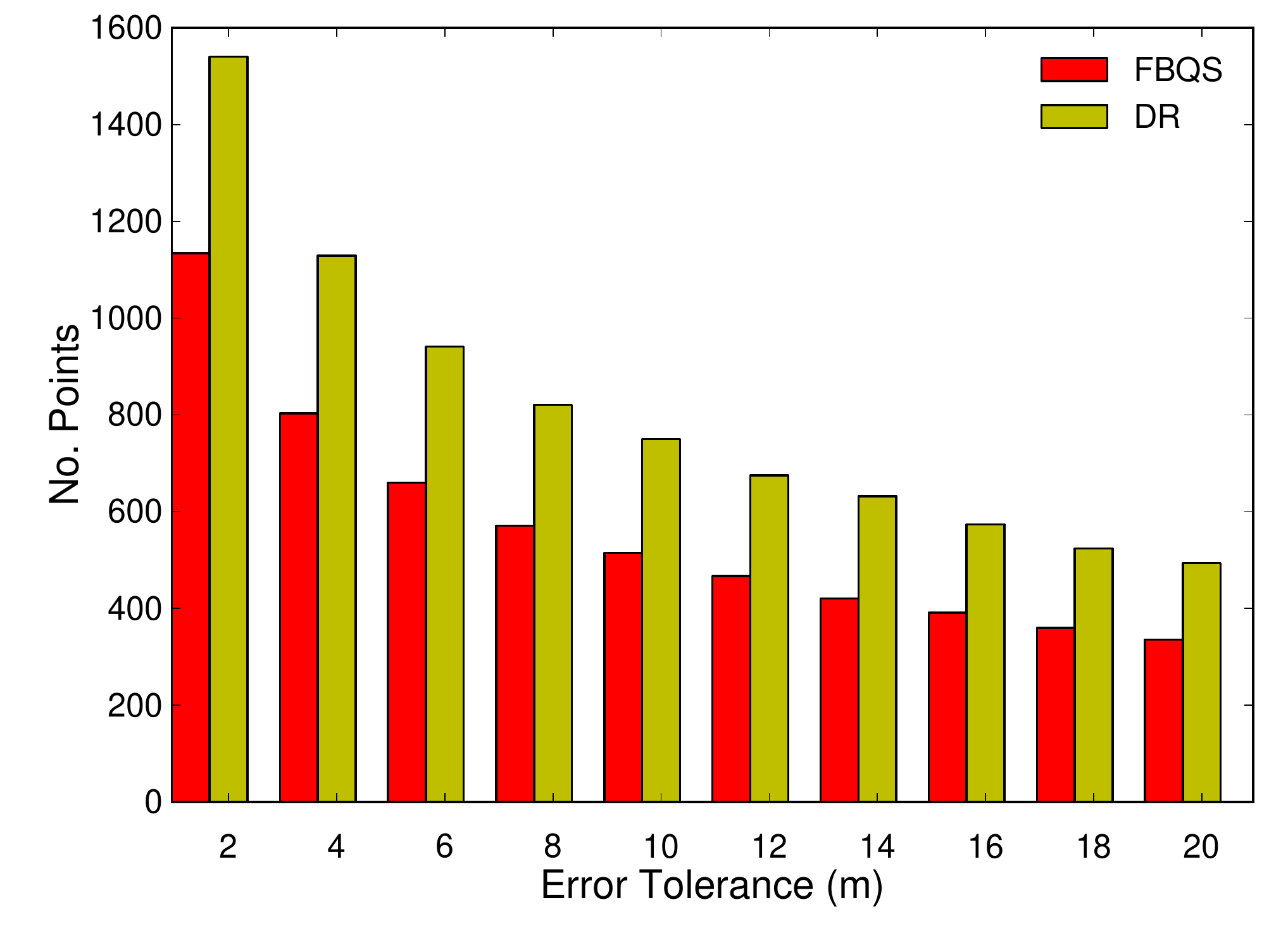}
}
\end{tabular}
\caption{Shape of Synthetic Data and Comparison of Number of Points
  Used (The lower the better)}
\vspace{-0.6cm}
\end{figure}

\subsubsection{Comparison with Dead Reckoning on Synthetic Data}
In Figure \ref{fig:syn} we show the simulated trajectories from our
statistical model. Visibly the trajectories show little physical
constraint and considerable variety in heading and turning angles. On
this dataset we study the performance comparison because on this
dataset we are able to simulate the tracking node
closely with high frequency sampling. Hence FBQS is used as 
a light-weight setup to fit such online environment.

We show in Figure \ref{fig:nps} the numbers of points taken after the
compression of $30,000$ points under different error tolerances. With
smaller $\epsilon$ such as $2~m$, DR uses $1,550$ points compared to
$1,100$ for FBQS, indicating that DR needs $40\%$ more points. As $\epsilon$ grows, DR's
performance tends to slowly approach FBQS in absolute numbers yet the
difference ratio becomes more significant. 
At $20~m$ error tolerance, FBQS
only takes $330$ points while DR uses $500$ points, the
difference ratio to FBQS is around $50\%$. 

Evidently, FBQS has achieved excellent 
compression rate compared to other existing online algorithms.

\subsubsection{Effect on Operational Time of Tracking Device}
Next we investigate how different online algorithms affect the maximum
operational time of the targeted device. This operational time
indicates how long the device can keep records of the locations 
before offloading to a server, without data loss.

In the real-life application, the nodes also store other sensor information such as acceleration, heading, temperature,
humidity, energy profiling, sampled at much higher frequencies
due to their relatively low energy cost. We assume that of the 1MBytes
storage, GPS traces can use up to 50KBytes, and that the sampling
rate of GPS is 1 sample per minute. Each GPS sample requires
at least 12 bytes storage (latitude, longitude, timestamp). For the
error tolerance, we use 10 meters as it is reasonable for both
animal tracking and vehicle tracking. The average compression rate at 10
meters for both datasets is used for the algorithms. For the DR
algorithm, we assume it uses 39\% more points than FBQS as shown in
Figure \ref{fig:nps} at the same tolerance.

Given the set up, the compression rate and
estimated operational time without data loss for each algorithm are
listed in Table \ref{tbl:eot}. We can see a maximum 36\% improvement
from FBQS over the existing methods (60 v.s. 44), and a maximum 41\% improvement
from BQS (62 v.s. 44).

\begin{table}[htp]
\centering
\caption{Estimated Operational Time}
\label{tbl:eot}
\begin{tabular}{ | c |c|c|c|c|c| }
\hline  & \textbf{BQS} & \textbf{FBQS} & \textbf{BDP}& \textbf{BGD}&
\textbf{DR}\\ 
\hline \textbf{Compression rate} & 4.8\% & 5.0\% & 6.65\%& 6.75\%& 6.65\%\\
\hline \textbf{Time (days)} & 62&60&45&44&45\\
\hline
\end{tabular}
\end{table}

\subsubsection{Run Time Efficiency}
We compared the run time efficiency of FBQS, BDP
and BGD. 87,704 points from the empirical traces are used as the test
data. The error tolerance is set to 10 meters. For BDP and BGD, to
minimize the effect of the buffer size, we report their performances
with different buffer sizes, as in Table \ref{tbl:ct}:
  
\begin{table}[htp]
\centering
\caption{Performance Comparison with different buffer sizes}
\label{tbl:ct}
\begin{tabular}{ | c |c|c|c|c|c| }
\hline \textbf{Buffer size (points)} && 32 & 64& 128& 256 \\ \hline
\multirow{3}{*}{\textbf{Compression rate}} 
& \textbf{FBQS} & 3.6\%&--- &--- &--- \\
& \textbf{BDP} & 6.8\%& 6.7\%& 5.4\%& 4.9\% \\
& \textbf{BGD} &6\%& 4.8\%& 4.6\%&4.4\%\\ \hline
\multirow{3}{*}{\textbf{Run time (ms)}} 
& \textbf{FBQS} &99&--- &--- &--- \\
& \textbf{BDP} & 76& 101& 163& 292 \\
& \textbf{BGD} &182&285&446&628\\
\hline
\end{tabular}
\end{table}

There are two notice-able advantages of FBQS from the
comparison. Firstly, both the compression rate and the run time efficiency of 
FBQS algorithm are stable, independent of the buffer size setting. 
Secondly, it offers competitive run time efficiency while providing
leading compression rate. The only case when BDP is able to
outperform FBQS in run time efficiency is when the buffer is set to
32, where BDP has a far worse compression rate (89\% more points).

%% file: main.bbl
\begin{thebibliography}{10}
\providecommand{\url}[1]{#1}
\csname url@samestyle\endcsname
\providecommand{\newblock}{\relax}
\providecommand{\bibinfo}[2]{#2}
\providecommand{\BIBentrySTDinterwordspacing}{\spaceskip=0pt\relax}
\providecommand{\BIBentryALTinterwordstretchfactor}{4}
\providecommand{\BIBentryALTinterwordspacing}{\spaceskip=\fontdimen2\font plus
\BIBentryALTinterwordstretchfactor\fontdimen3\font minus
  \fontdimen4\font\relax}
\providecommand{\BIBforeignlanguage}[2]{{%
\expandafter\ifx\csname l@#1\endcsname\relax
\typeout{** WARNING: IEEEtran.bst: No hyphenation pattern has been}%
\typeout{** loaded for the language `#1'. Using the pattern for}%
\typeout{** the default language instead.}%
\else
\language=\csname l@#1\endcsname
\fi
#2}}
\providecommand{\BIBdecl}{\relax}
\BIBdecl

\bibitem{Jurdak_tosn13}
R.~Jurdak, P.~Corke, A.~Cotillon, D.~Dharman, C.~Crossman, and G.~Salagnac,
  ``Energy-efficient localisation: Gps duty cycling with radio ranging,''
  \emph{Transactions on Sensor Networks}, vol.~9, no.~2, 2013.

\bibitem{van2010survey}
D.~Van~Krevelen and R.~Poelman, ``A survey of augmented reality technologies,
  applications and limitations,'' \emph{International Journal of Virtual
  Reality}, vol.~9, no.~2, p.~1, 2010.

\bibitem{DBLP:journals/tosn/EisenmanMLPAC09}
S.~B. Eisenman, E.~Miluzzo, N.~D. Lane, R.~A. Peterson, G.-S. Ahn, and A.~T.
  Campbell, ``Bikenet: A mobile sensing system for cyclist experience
  mapping,'' \emph{TOSN}, vol.~6, no.~1, 2009.

\bibitem{DBLP:conf/ipsn/JurdakSKKCMW13}
R.~Jurdak, P.~Sommer, B.~Kusy, N.~Kottege, C.~Crossman, A.~Mckeown, and
  D.~Westcott, ``Camazotz: multimodal activity-based gps sampling,'' in
  \emph{IPSN}, 2013, pp. 67--78.

\bibitem{nagy2010hierarchical}
M.~Nagy, Z.~{\'A}kos, D.~Biro, and T.~Vicsek, ``Hierarchical group dynamics in
  pigeon flocks,'' \emph{Nature}, vol. 464, no. 7290, pp. 890--893, 2010.

\bibitem{douglas_peucker}
D.~H. Douglas and T.~K. Peucker, \emph{Algorithms for the Reduction of the
  Number of Points Required to Represent a Digitized Line or its
  Caricature}.\hskip 1em plus 0.5em minus 0.4em\relax John Wiley and Sons, Ltd,
  2011, pp. 15--28.

\bibitem{Hershberger92speedingup}
J.~Hershberger and J.~Snoeyink, ``Speeding up the douglas-peucker
  line-simplification algorithm,'' in \emph{Proc. 5th Intl. Symp. on Spatial
  Data Handling}, 1992, pp. 134--143.

\bibitem{squishe}
J.~Muckell, J.~Olsen, PaulW., J.-H. Hwang, C.~Lawson, and S.~Ravi,
  ``\BIBforeignlanguage{English}{Compression of trajectory data: a
  comprehensive evaluation and new approach},''
  \emph{\BIBforeignlanguage{English}{GeoInformatica}}, pp. 1--26, 2013.

\bibitem{opening_window}
E.~Keogh, S.~Chu, D.~Hart, and M.~Pazzani, ``An online algorithm for segmenting
  time series,'' in \emph{Proc. ICDM 2001}, 2001, pp. 289--296.

\bibitem{ChenXF12}
M.~Chen, M.~Xu, and P.~Fränti, ``A fast $o(n)$ multiresolution polygonal
  approximation algorithm for gps trajectory simplification.'' \emph{IEEE
  Transactions on Image Processing}, vol.~21, no.~5, pp. 2770--2785, 2012.

\bibitem{Long_2013}
C.~Long, R.~C.-W. Wong, and H.~V. Jagadish, ``Direction-preserving trajectory
  simplification,'' \emph{Proc. VLDB Endow.}, vol.~6, no.~10, pp. 949--960,
  Aug. 2013.

\bibitem{squish}
J.~Muckell, J.-H. Hwang, V.~Patil, C.~T. Lawson, F.~Ping, and S.~S. Ravi,
  ``Squish: An online approach for gps trajectory compression,'' in \emph{Proc.
  COM.Geo '11}.\hskip 1em plus 0.5em minus 0.4em\relax ACM, 2011, pp.
  13:1--13:8.

\bibitem{sttrace}
M.~Potamias, K.~Patroumpas, and T.~Sellis, ``Sampling trajectory streams with
  spatiotemporal criteria,'' in \emph{Scientific and Statistical Database
  Management, 2006. 18th International Conference on}, 2006, pp. 275--284.

\bibitem{liu_MBR}
G.~Liu, M.~Iwai, and K.~Sezaki, ``An online method for trajectory
  simplification under uncertainty of gps,'' in \emph{Information and Media
  Technologies; VOL.8; NO.3}, 2013, pp. 665--674.

\bibitem{Trajcevski06on-linedata}
G.~Trajcevski, H.~Cao, P.~Scheuermanny, O.~Wolfsonz, and D.~Vaccaro, ``On-line
  data reduction and the quality of history in moving objects databases,'' in
  \emph{Proc. MobiDE '06}.\hskip 1em plus 0.5em minus 0.4em\relax ACM, 2006,
  pp. 19--26.

\bibitem{Kjaergaard:2009:EER:1555816.1555839}
M.~B. Kj{\ae}rgaard, J.~Langdal, T.~Godsk, and T.~Toftkj{\ae}r, ``Entracked:
  Energy-efficient robust position tracking for mobile devices,'' in
  \emph{Proc. MobiSys '09}.\hskip 1em plus 0.5em minus 0.4em\relax ACM, 2009,
  pp. 221--234.

\bibitem{Heckbert95surveyof}
P.~S. Heckbert and M.~Garland, ``Survey of polygonal surface simplification
  algorithms,'' 1995.

\bibitem{Knuth:1997:ACP:270146}
D.~E. Knuth, \emph{The Art of Computer Programming, Volume 2 (3rd Ed.):
  Seminumerical Algorithms}.\hskip 1em plus 0.5em minus 0.4em\relax Boston, MA,
  USA: Addison-Wesley Longman Publishing Co., Inc., 1997.

\bibitem{citeulike:3939699}
J.~Hightower, ``{SpotON: An Indoor 3D Location Sensing Technology Based on RF
  Signal Strength}.''

\bibitem{Cao:2006:SDR:1147679.1147681}
H.~Cao, O.~Wolfson, and G.~Trajcevski, ``Spatio-temporal data reduction with
  deterministic error bounds,'' \emph{The VLDB Journal}, vol.~15, no.~3, pp.
  211--228, Sep. 2006.

\bibitem{Risken}
H.~Risken, \emph{{The Fokker-Planck Equation: Methods of Solutions and
  Applications}}, 2nd~ed., ser. Springer Series in Synergetics.\hskip 1em plus
  0.5em minus 0.4em\relax Springer, Sep. 1996.

\end{thebibliography}
